# Bridgman-grown (Cd,Mn)Te and (Cd,Mn)(Te,Se): A comparison of suitability for X and gamma detectors

Aneta Masłowska[1,*], Dominika M. Kochanowska[1], Adrian Sulich[1], Jaroslaw Z. Domagala[1], Marcin Dopierała[1], Michał Kochański[1], Michał Szot[1,2], Witold Chromiński[3], Andrzej Mycielski[1,4]

[1] Institute of Physics, Polish Academy of Sciences, Aleja Lotników 32/46, 02-668 Warsaw, Poland

[2] International Research Centre MagTop, Institute of Physics, Polish Academy of Sciences, Aleja Lotników 32/46, 02-668 Warsaw, Poland

[3] Warsaw University of Technology, Faculty of Materials Science and Engineering, Wołoska 141, 02-507 Warsaw, Poland

[4] Puremat Technologies Sp. z o.o., Aleja Lotników 32/46, 02-668 Warsaw, Poland

[*] Corresponding author. E-mail: wardak@ifpan.edu.pl

## Abstract

This study explores the suitability of semi-insulating compounds, specifically (Cd,Mn)Te and (Cd,Mn)(Te,Se), as materials for room temperature X-ray and gamma-ray detectors. These compounds were grown using the Bridgman method, known for its efficient growth rate. The investigation aims to compare their crystal structure, mechanical properties, optical characteristics, and radiation detection capabilities.

The addition of selenium to (Cd,Mn)Te increased the compound's hardness. However, (Cd,Mn)(Te,Se) exhibited one order of magnitude higher etch pit density compared to (Cd,Mn)Te. Photoluminescence analysis at low temperatures revealed the presence of defect states in both materials, characterized by shallow and deep donor-acceptor pair transitions (DAP). Annealing in cadmium vapors effectively eliminated DAP luminescence in (Cd,Mn)Te but not in (Cd,Mn)(Te,Se).

Spectroscopic performance assessments indicated that the (Cd,Mn)Te detector outperformed the (Cd,Mn)(Te,Se) detector in responding to a Co-57 source. The reduced performance in the latter case may be attributed to either the presence of a deep trap related to deep DAP luminescence, minimally affected by annealing, or the dominant presence of block-like structures in the samples, as indicated by X-ray diffraction measurements. The block-like structures in (Cd,Mn)(Te,Se) showed ten times larger misorientation angles compared to the (Cd,Mn)Te crystals. (Cd,Mn)Te crystal revealed excellent single crystal properties, demonstrated by narrower omega scan widths. The study also highlights the influence of grain boundaries and twins on crystal structure quality.

In our opinion, Bridgman-grown (Cd,Mn)Te shows greater promise as a material for X-ray and gamma-ray detectors compared to (Cd,Mn)(Te,Se).

**Keywords**

Annealing, Crystal Growth, Photoluminescence, X-ray and Gamma-ray Detectors, XRD Mapping.

## 1. Introduction

CdTe-based compounds, such as (Cd,Zn)Te [1–4], (Cd,Zn)(Te,Se) [5–17], (Cd,Mn)Te [18–23], (Cd,Mn)(Te,Se) [24,25], (Cd,Mg)Te [26–31], Cd(Te,Se) [32–38] are currently being tested and as materials for room-temperature X- and gamma-ray detectors. In these nuclear detectors, it is crucial to maximize the flow of radiation-induced charge carriers through the detector volume to the respective electrodes. Therefore, it is highly desirable for the material to have minimal defect concentrations to ensure optimal detector performance. In practice, this translates to the requirement of high resistivity ($\rho \sim 10^9$-$10^{10}$ $\Omega$ cm) and a high mobility-lifetime product ($\mu\tau \sim 10^{-3}$ cm$^2$V$^{-1}$) for the crystals [39].

In our laboratory, we regularly conduct investigations on high resistivity (Cd,Mn)Te, which material did we propose for X-ray and gamma detectors [40] and which has proven to be a promising X-ray and gamma radiation detector [41]. Furthermore, for crystal growth, we employ the Bridgman method, enabling the relatively fast production of large (≥ 1.5 in.) crystals with a crystal growth rate of several mm per hour, as compared to methods such as the traveling heater method (THM), where the crystal growth rate is 3-5 mm per day [42,43]. Our motivation to study compounds alloyed with selenium, specifically (Cd,Mn)(Te,Se), stemmed from the literature findings. In the field of X-ray and gamma-ray detectors, there is currently a heated debate regarding the use of selenium-alloyed CdTe crystals. In this article, we present new data and express our position on this matter.

According to previous research on (Cd,Zn)Te and (Cd,Zn)(Te,Se) [8,44], the presence of selenium plays a crucial role in inhibiting the development of sub-grain boundary networks, and increases the hardness of the quaternary material (Cd,Zn)(Te,Se). Additionally, selenium significantly reduces the concentration of tellurium inclusions [43]. Subsequently, it can be inferred from UV-Vis absorbance data that changes in the bandgap values along the growth direction (reduced length of 0.1 to 0.9 of the ingot) in $Cd_{0.9}Zn_{0.1}Te$ crystals amount to 31 meV, whereas in $Cd_{0.9}Zn_{0.1}Te_{0.98}Se_{0.02}$, they are only 21 meV [45]. This indicates that in (Cd,Zn)(Te,Se) crystals, a more uniform difference of bandgap values along the growth direction is achieved due to reduced segregation effects. A homogeneous distribution of the energy bandgap in crystals is important because it influences the distribution of other properties, such as resistivity and absorption edge [39].

In this work, instead of the more commonly investigated (Cd,Zn)Te, we utilize (Cd,Mn)Te due to several advantages of Mn-alloyed CdTe over Zn-alloying. Firstly, the segregation coefficient of Mn in CdTe is close to unity, specifically 0.95 [46], whereas the segregation coefficient of Zn in CdTe is 1.3 [47]. This results in a more homogeneous distribution of Mn in CdTe, whereas in (Cd,Zn)Te, there is a higher concentration of Zn at the first-to-freeze part of the ingot. However, there are reports that the addition of Se to (Cd,Zn)Te reduces Zn segregation in the ingot, and a homogeneous chemical composition can be achieved in the axial and radial directions in 90% of the volume of (Cd,Zn)(Te,Se) crystals obtained by the THM method [43]. Secondly, for a detector to operate at room temperature, the semiconductor material should have an appropriate energy bandgap, ranging from 1.5 to 2.2 eV [39]. Achieving the desired energy bandgap involves adding a larger quantity of Zn to CdTe in comparison to introducing Mn to CdTe. This is because the energy bandgap of (Cd,Zn)Te undergoes a slower transformation with the addition of Zn, changing at a rate of 6.7 meV per atomic percent of Zn [48]. Conversely, introducing Mn to CdTe alters the energy bandgap by 13 meV for every atomic percent of Mn [49]. Furthermore, the addition of Se to CdTe reduces the energy bandgap in crystals typically chosen for X-ray and gamma-ray detector applications, specifically those containing less than or equal to 2% Se. In $CdTe_{1-x}Se_x$ crystals, the energy gap decreases as x increases, at a rate of approximately 4 meV per atomic percent of Se for $x \leq 0.1$. This rate decreases for $0.1 \leq x \leq 0.4$, reaches its minimum at $x \approx 0.4$, and then begins to increase as x continues to rise [50]. This consideration is significant. Restricting the quantity of the third or fourth alloying component helps limit the formation of new defects within the crystal structure. Thirdly, it has been experimentally observed that larger grains can be obtained in (Cd,Mn)Te than in (Cd,Zn)Te, enabling the production of larger volume detector plates from such a crystal.

In this study, we investigate two compounds grown by the Bridgman method: (Cd,Mn)Te and (Cd,Mn)(Te,Se), and we compare their suitability for X-ray and gamma-ray detectors. The research involves examining the hardness and employing the etch pit density technique to determine if harder materials display fewer detrimental sub-grain boundaries and their networks. We conducted microstructure imaging using infrared microscopy. Subsequently, we conduct a detailed characterization of the crystalline structure of selected large-surface-area single crystal samples, such as $30 \times 30$ mm$^2$. We draw lattice constant and omega scan maps for these compounds. The photoluminescence of the as-grown samples is investigated, and by comparing the results with samples annealed in Cd and Se vapors, we discuss the presence of defects in the as-grown crystals. Finally, we present the energy spectrum from a Co-57 source recorded using our selected and optimized crystal.

## 2. Materials and methods

Crystals of $Cd_{1-x}Mn_xTe$ and $Cd_{1-x}Mn_xTe_{1-y}Se_y$ were grown using the low-pressure Bridgman method, with x set at either 5% or 7%, and y at 2%. The growth process employed high-purity materials, specifically 7N Cd, 7N Te, 6N Mn, and 6N Se. These crystals had diameters of either 2 or 3 in. Hardness investigations were carried out on a $Cd_{0.95}Mn_{0.05}Te_{0.98}Se_{0.02}$ crystal as the primary sample, while $Cd_{0.93}Mn_{0.07}Te_{0.98}Se_{0.02}$ crystals were utilized in other measurements. In the context of (Cd,Mn)Te crystals, those with a 5% Mn composition were consistently examined. The crystal growth process was performed under Te-rich conditions, involving the addition of 30 mg to 100 mg of extra Te per 100 g of material. For compensation, a co-doping of indium or vanadium was used. The concentration of indium was $1 \times 10^{17}$ cm$^{-3}$ (for $Cd_{0.95}Mn_{0.05}Te$), or $5 \times 10^{16}$ cm$^{-3}$ (for $Cd_{0.95}Mn_{0.05}Te_{0.98}Se_{0.02}$), or $1 \times 10^{14}$ cm$^{-3}$ (for $Cd_{0.93}Mn_{0.07}Te_{0.98}Se_{0.02}$), while that of vanadium was $1\times10^{13}$ $cm^{-3}$ for all crystals. To visualize the grain boundaries in the crystals, they were etched using the Durose solution [51]. Next, the samples were mechano-chemically polished using a 2% bromine solution in methanol and ethylene glycol.

Annealing processes were carried out for 168 hours in vacuum-sealed quartz ampoules in Cd vapors at the temperature of 800 °C or in Se vapors at 350 °C.

Mechanical properties were tested by measuring the microhardness of the polished samples on the cadmium side, i.e. (111)A plane, using a Vickers indenter with a load of 50 g for 15 s. For the purpose of comparison, one CdTe crystal, one Cd(Te,Se) crystal with a 5% Se composition, and two (Cd,Zn)Te crystals with 5% and 12% Zn, respectively, were also utilized in these studies.

The etch pit density was examined using the E-Ag1 Inoue solution, which consists of 0.5 parts $AgNO_3$ and 10 parts solution E. Solution E is composed of 5 parts $HNO_3$, 10 parts $H_2O$, and 2 parts $K_2Cr_2O_7$ [52]. The etching time was equal to 90 s. The (111)A side was observed. For microscopic observations, an Olympus BX51 microscope equipped with an Olympus XC10 CCD camera was utilized, either in reflection or infrared transmission mode, depending on the purpose.

In the diffraction studies, a high-resolution Philips X'Pert MRD diffractometer was employed, featuring a monochromatized source of $CuK_{\alpha1}$ radiation ($\lambda$ = 1.5406 Å) and further equipped with a homemade mask and slit. This homemade mask was responsible for reducing the dimensions of the X-ray beam to $0.5\times1.0$ $mm^2$, enabling the collection of diffraction data from the samples point by point and line by line. X-ray scans were performed on sample areas measuring $18\times20$ $mm^2$. Bragg angle measurements were carried out, with a focus on the 333 reflex, to generate maps of lattice constant variations within the samples. Omega curve measurements, depicting the intensity of the diffracted beam on the crystal as a function of the omega angle, $\omega$, which is the angle between incident X-ray beam and the surface of the sample, were conducted to create delta omega maps and full width at half maximum (FWHM) maps. Both Bragg angle and omega curve measurements were carried out with a step size of 1 mm for X and 2 mm for Y.

Omega curve measurements were conducted in either double-axis mode (DA) or triple-axis mode (TA). In the TA mode, an analyzer was utilized to enhance the resolution, reducing it from 18 arcsec (DA) to 8 arcsec (TA), as determined by a measurement of a Si(111) reference sample. In the DA mode the detector was fully opened. The omega angle measured with the analyzer is referred to as $\omega_{TA}$, while without the analyzer, it is denoted as $\omega_{DA}$.

The omega angle was determined at the peak of the curve or, in cases of multiple peaks, at the extreme peaks. Delta omega, $\Delta\omega$, was then calculated by subtracting these values from each other. In instances where only one peak was present on the curve, it was assumed that delta omega equaled zero. This data was utilized to generate delta omega maps.

The photoluminescence (PL) studies were carried out on the cleaved samples and the excitation energy was equal to 2.33 eV. The PL spectra were obtained at 5 K in a continuous flow cryostat with a photomultiplier.

The spectroscopic response at room temperature of the pixelated detectors was checked using a Co-57 source and the Spectroscopic Pixel Mapping machine by Eurorad. For this purpose, electrical contacts, made of a gold-palladium alloy in an 80:20 ratio, were deposited using the Quorum Sputter Coater Q150T. Pixels on the anode, which crystallographically represents the cadmium side, were produced using the photolithography method. The cathode remained planar. The detector received radiation from the cathode side and was maintained at a bias voltage of −400 V. The shaping time was set to 1 μs.

## 3. Results and discussion

In our laboratory, by properly doping with In, V, or both In and V, we can obtain (Cd,Mn)Te and (Cd,Mn)(Te,Se) crystals with average resistivity on the order of $10^9$ Ω cm and a mobility-lifetime product on the order of $10^{-3}$ $cm^2V^{-1}$ [41]. We are also capable of producing large single crystal plates, as depicted in Fig. 1. Fig. 1a shows a sliced (Cd,Mn)Te plate, which was cut perpendicular to the growth axis from a 2-in. ingot, ground, and then etched with Durose solution to reveal grain boundaries. It is evident that the plate is monocrystalline. On the other hand, Fig. 1b shows a (111)-oriented polished plate that has been prepared for detector fabrication, i.e., for the application of electrical contacts. This plate is also monocrystalline and has large dimensions of approximately 30×30 $mm^2$. Visual examination of the obtained crystals yielded satisfactory results. These results provide an excellent basis for further research, as the field of room-temperature X-ray and gamma semiconductor detectors demands high-resistivity ($>10^9$ Ω cm), defect-free, monocrystalline, large (more than 5-8 $cm^2$), and sufficiently thick (>3 mm) plates for effective operation, i.e., to ensure effective interactions between high-energy radiation and the detector material. In Subsection 3.2, we will present more advanced studies on the crystal structure.

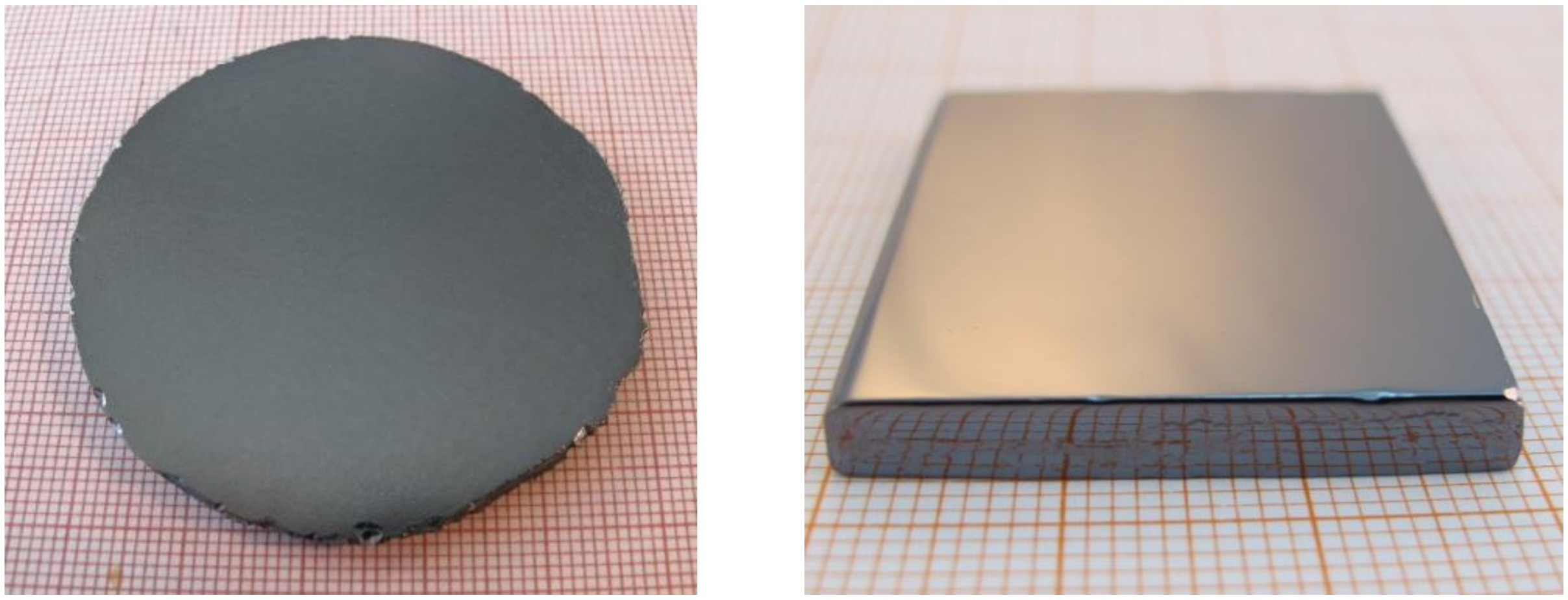

(a) (b)

Fig. 1. The image of the single crystal (Cd,Mn)Te samples. (a) Crystal plate cut from the 2-in. ingot and etched with Durose solution. No grain boundaries or twins are visible. (b) (111)-oriented polished monocrystalline (Cd,Mn)Te plate of a specified shape for a detector. It was cut from a 3-in. ingot.

### 3.1 Hardness

The influence of Mn and Se additives on the hardness of the formed CdTe-based compounds was examined. As shown in Fig. 2, a five percent addition of manganese or selenium to CdTe increases the hardness of the compound, with the influence of selenium being stronger. The hardness of CdTe alloyed with manganese (5%) and selenium (2%), i.e., (Cd,Mn)(Te,Se) compound, falls between the hardness of (Cd,Mn)Te (Mn 5%) and Cd(Te,Se) (Se 5%). For comparison, we also investigated two crystals of (Cd,Zn)Te with different Zn contents, namely 5% and 12%. The crystals with Zn exhibited the highest hardness among all the samples examined, and a higher Zn content resulted in increased hardness, which is consistent with the literature data [53]. In the comparison, we included the hardness result for CdTe crystal both alloyed with Zn and Se, measured by other authors [11]. The addition of 2% Se to (Cd,Zn)Te further hardened the material. The result maintains the same trend. Despite the greater hardness of (Cd,Zn)Te crystals, the focus of this study lies on (Cd,Mn)Te crystals because, as mentioned in Section 1, it is easier to attain larger grains in the latter.

The addition of some additional elements to the CdTe matrix, like Mn, Se, or Zn, alters its structure. These changes can make it more challenging for atoms and dislocations to move within the lattice, resulting in material hardening. Additionally, by introducing an additional element to the CdTe matrix, the bond lengths, like Cd-Te, Mn-Te, Zn-Te, Cd-Se, are changed, leading to the formation of strong and stable atomic connections [54]. These bonds can hinder atom movement and impede material deformation.

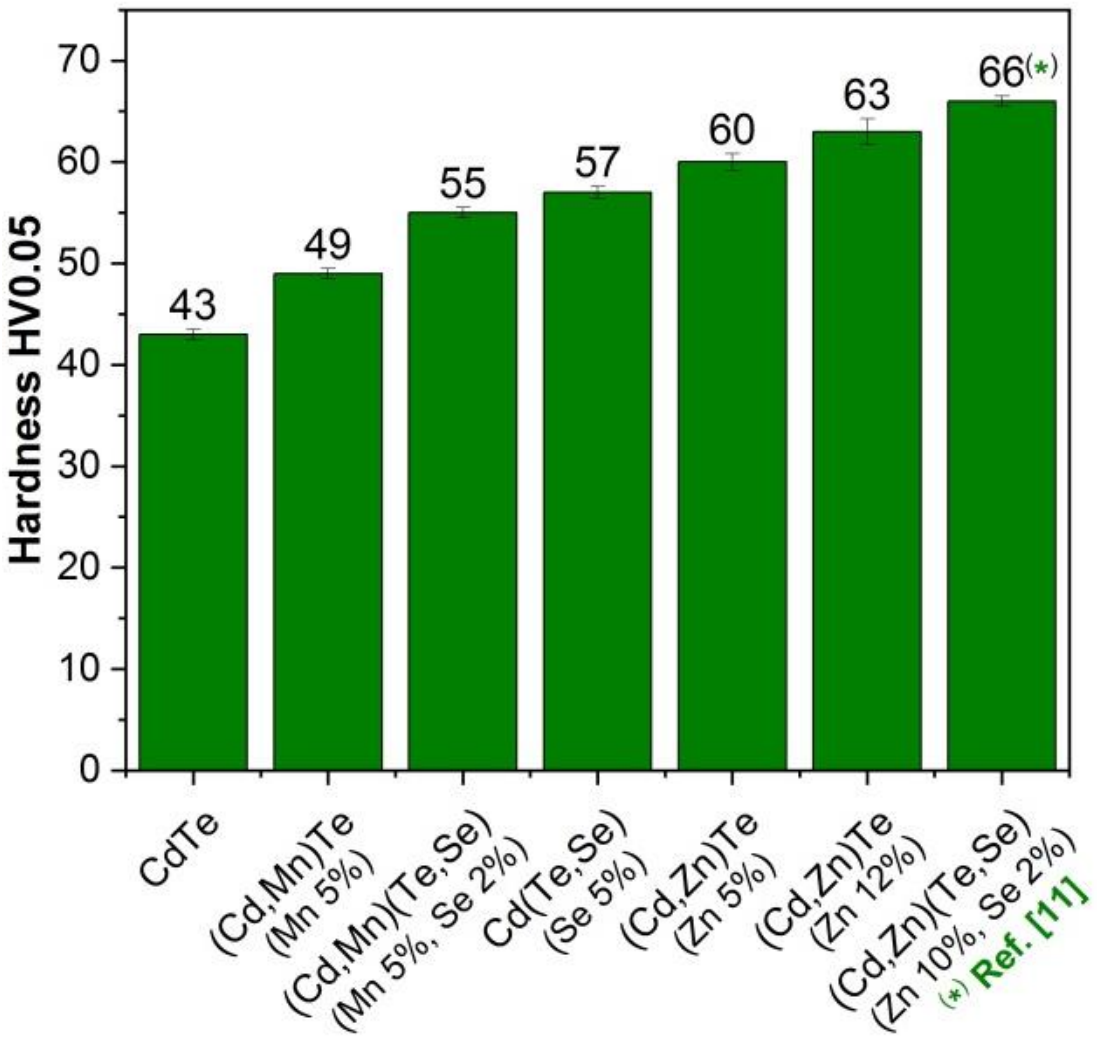

Fig. 2. Hardness of the selected CdTe-based compounds measured on the (111)A cadmium side.

### 3.2 Crystal quality

#### 3.2.1 Etch pit density

A test was conducted to examine whether higher hardness values translate into a smaller population of sub-grain boundaries, which are often encountered in CdTe-based compounds produced using the Bridgman method and pose a significant issue in detector performance. Typically, sub-grain boundaries are investigated using the White Beam X-ray Diffraction Tomography method [8,55–59]. We employed an etching method to reveal etch pits using the Inoue solution, as the Nakagawa solution [60] yielded no results on our samples. Sub-grain boundaries, which have a small misorientation angle, are formed by dislocation clusters. Sub-grain sizes are on the order of hundreds of micrometers [56]. Hence, if we observe any clusters of etch pits, which form on the crystal surface at the location where dislocations initiate, it could suggest the presence of sub-grain boundaries in the investigated crystal. The formation of etch pits related to dislocations occurs due to the interplay between the stress field caused by dislocations and the surface energy [61].

Fig. 3 illustrates microscopic images of the ~(111)A surface of three investigated by us compounds, which were etched with the Inoue solution to visualize etch pits. Fig. 3a and 3d depict the surface of a CdTe reference sample at different magnifications. Fig. 3b and 3e show (Cd,Mn)Te, Fig. 3c and 3f represent (Cd,Mn)(Te,Se). The straight lines in Fig. 3c represent scratches on the sample surface. The CdTe and (Cd,Mn)(Te,Se) samples exhibit a high density of etch pits, on the order of $10^5$ $cm^{-2}$. However, in the CdTe sample, the average size of etch pits is significantly larger compared to (Cd,Mn)(Te,Se), measuring approximately 40 μm and 5 μm, respectively. The density of etch pits is the lowest in the (Cd,Mn)Te sample, namely $10^4$ $cm^{-2}$, and their size ranges between 3 and 5 μm. The larger size of the etch pits in CdTe than in (Cd,Mn)Te and (Cd,Mn)(Te,Se) may be attributed to the larger stress fields generated by dislocations in that region. Observations of the etch pits revealed their uniform distribution, without the formation of clusters resembling small-angle boundaries. This suggests the absence of sub-grain boundaries in each of the investigated compounds, although this cannot be conclusively determined by this method. However, it is certain that the (Cd,Mn)Te sample exhibited the smallest dislocation density on its surface and demonstrated the best quality among the samples examined.

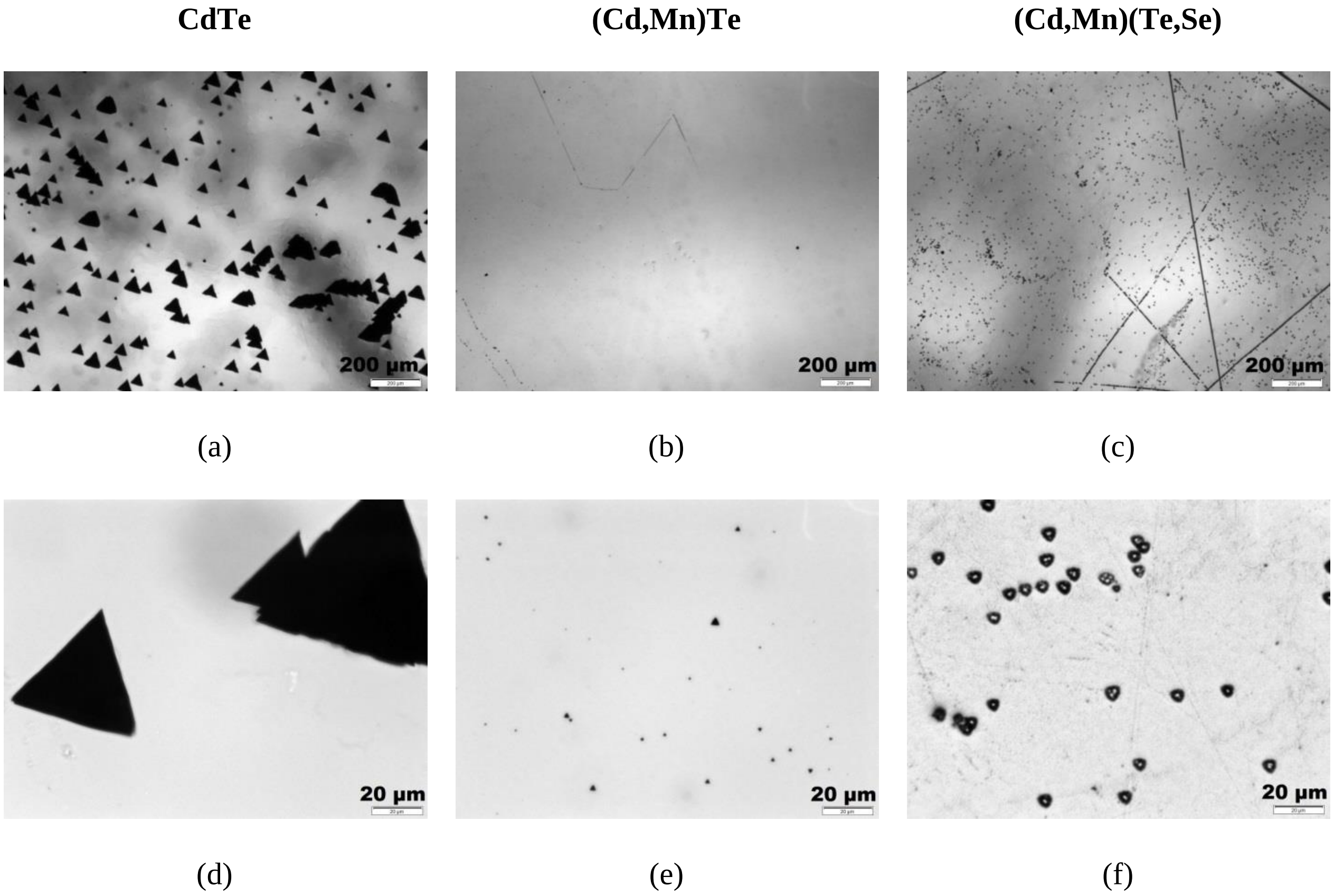

Fig. 3. Etch Pit Density (EPD) images of CdTe – (a) and (d), $Cd_{0.95}Mn_{0.05}Te$ – (b) and (e), and $Cd_{0.93}Mn_{0.07}Te_{0.98}Se_{0.02}$ – (c) and (f). The tests were performed using the Inoue etchant on the ~(111)A plane. Figs. 3 d, e, and f are magnified 10 times compared to Figs. 3 a, b, and c, respectively.

### 3.2.2 Lattice constant

We conducted a lattice constant mapping on monocrystalline samples of (Cd,Mn)Te (Fig. 4a) and (Cd,Mn)(Te,Se) (Fig. 4b). The lattice constant changes, denoted in Fig. 4 as Δa/<a>, are expressed as in Eq. 1:

$$\frac{\Delta a}{\langle a \rangle} = \frac{a - \langle a \rangle}{\langle a \rangle} \cdot 10^6 \quad [ppm] \qquad (1)$$

where: a – local value of lattice constant, <a> – the arithmetic mean value of all local values a determined at different locations along the sample.

The average value of lattice constant for $Cd_{0.95}Mn_{0.05}Te$ is 6.47658 Å, and for $Cd_{0.93}Mn_{0.07}Te_{0.98}Se_{0.02}$ is 6.46411 Å, both with a standard deviation of 0.00008 Å. In Fig. 4, the deviations from the average lattice constant value along the sample are depicted as very small, on the order of parts per million (ppm). This indicates that both crystals exhibit a high level of uniformity in terms of lattice constant distribution. The results provide a very solid foundation for further research on these crystals.

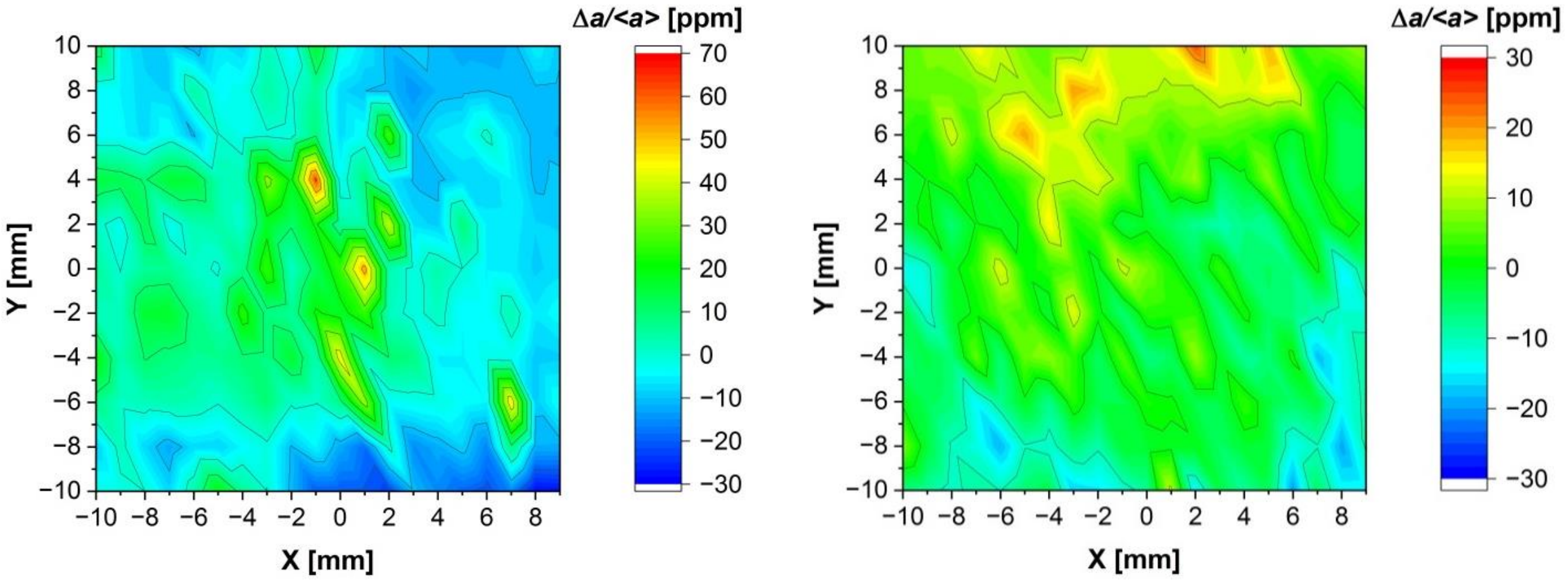

(a) (b)

Fig. 4. Lattice constant maps. (a) $Cd_{0.95}Mn_{0.05}Te$; (b) $Cd_{0.93}Mn_{0.07}Te_{0.98}Se_{0.02}$.

### 3.2.3 Presence of blocks/grains and their mutual misorientation

Delta omega maps were prepared to check the presence of blocks/grains in our samples and, if they were present, to determine their mutual misorientation. Delta omega map of the (Cd,Mn)Te sample is depicted in Fig. 5a. A non-zero delta omega value indicates the presence of blocks (consisting of two or more) that exhibit misorientation relative to each other, with the delta omega value representing the maximum misorientation between these blocks. Conversely, a delta omega value of zero signifies that, at that specific measurement point, only a single peak was recorded, indicating the absence of blocks or grains.

In Fig. 5a, a delta omega value of zero is found in the majority of measurement points. Out of the 220 points on the delta omega map, only 5 points show an omega scan curve with two peaks, indicating the presence of two misoriented blocks. This indicates that in 215 measurement points, which accounts for approximately 98% of the 18×22 $mm^2$ area of this (Cd,Mn)Te sample, a single peak was recorded, implying the absence of grains or blocks with different orientations. Consequently, it can be concluded that a well-defined monocrystal is observed within the resolution limits of our measurement method.

The map in Fig. 5b illustrates how the intensity (signal strength) of specific omega angles is varied in this (Cd,Mn)Te sample. This map was created using data from 20 omega curves collected at 20 points in the sample, along the Y = −8 line, with an X step of 1 mm. The aim of this was to visualize variations in the omega angle at different points (X, −8) within the sample. Here, the value of the omega angle should be determined based on the angle corresponding to the highest intensity, which signifies the peak of the signal. The choice of the Y = −8 line was motivated by the presence of several points with notably higher deviations, as clearly indicated in Fig. 5a, where the Y = −8 line is marked by a red dashed line.

In Fig. 5b, it is evident that the omega angle, depicted as points with the highest intensity, i.e., corresponding to the maximum of the omega curve, remains constant in the range from X = −10 to X = 0.5. At X = 0.5, a shift of the X-ray beam from one block (grain) to another is observed. Subsequently, from X = 2 to X = 9, the omega angle is once again maintained at a near-constant value. The maximum misorientation angle between these two blocks is measured at 50 arcsec. It can be observed that in the (Cd,Mn)Te sample, despite selecting a line along the Y-axis in the sample with poorer X-ray results for the creation of the map in Fig. 5b, the omega angles remain nearly constant.

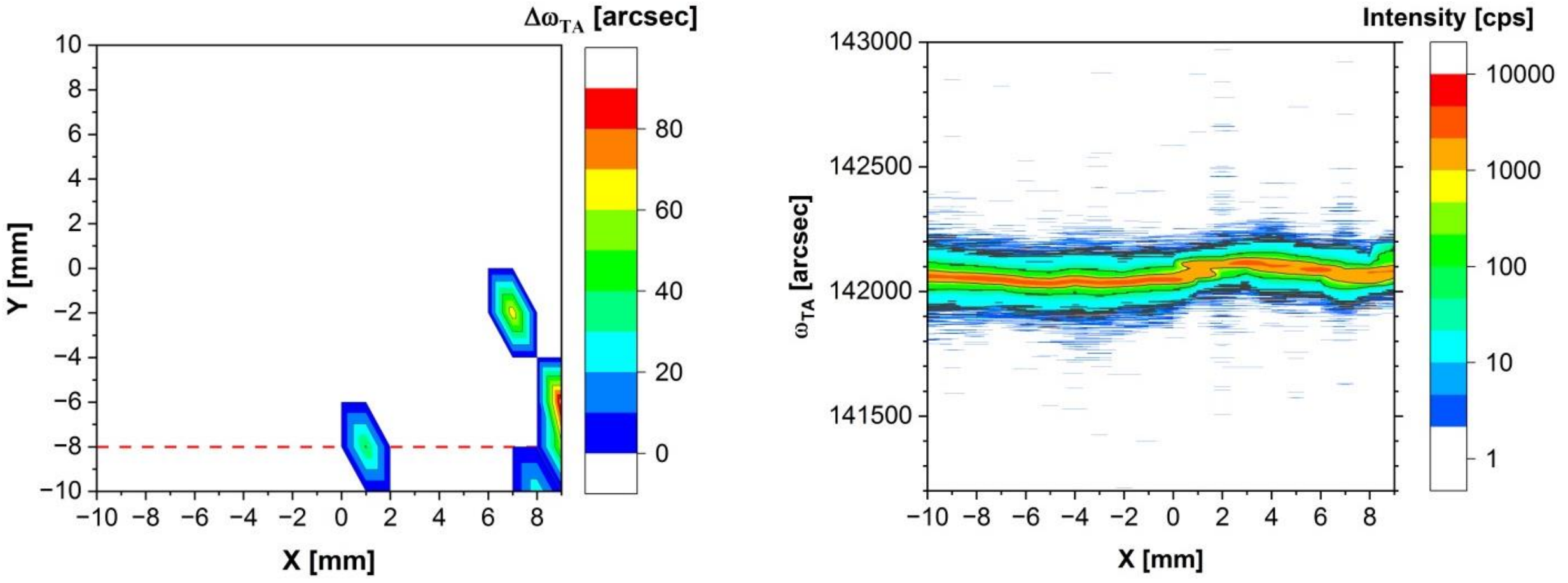

(a) (b)

Fig. 5. $Cd_{0.95}Mn_{0.05}Te$ results. (a) Delta omega map of $Cd_{0.95}Mn_{0.05}Te$ in the triple axis mode. (b) The map of the intensity of omega values, $\omega_{TA}$, obtained from 20 omega scans conducted along the Y = −8 line, with a 1 mm X step. The Y = −8 line is marked in Fig. 5a with a red dashed line.

Fig. 6a depicts the delta omega map of a (Cd,Mn)(Te,Se) sample. This map was also measured in triple axis mode. In the case of the (Cd,Mn)(Te,Se) crystal, a higher number of measurement points with non-zero values of delta omega are observed, indicating the presence of multiple blocks or grains. Furthermore, the maximum misorientation between these blocks, represented by the delta omega value, is significantly larger compared to (Cd,Mn)Te, on the order of 100 arcsec (with a maximum of 800 arcsec), whereas for (Cd,Mn)Te, it was on the order of 10 arcsec (with a maximum of 90 arcsec).

The changes of the intensity of omega angle values in the omega scans of a (Cd,Mn)(Te,Se) sample conducted along the Y = −10 line are depicted in Fig. 6b. In the case of the (Cd,Mn)(Te,Se) crystal, similar to (Cd,Mn)Te (Fig. 5b), the map was generated based on 20 omega scans along the Y-axis, and a Y line was chosen where a greater number of measurement points with higher (worse) delta omega values were encountered. Here, a significant dispersion of omega angle values is evident.Fig. 6b is on the same angular scale as Fig. 5b. The variation in these values between the red areas from Fig. 6b, those with the highest intensity, is 720-1080 arcsec (0.2-0.3 degrees). Let us recall that in the worst location of the (Cd,Mn)Te sample, Y = −8, variations in the omega angle were at the level of 50 arcsec (Fig. 5b). Although the (Cd,Mn)(Te,Se) sample etched with Durose's solution appeared to be monocrystalline to the naked eye, X-ray studies revealed the presence of misoriented blocks within it. This is clearly visible in Fig. 6a, where monocrystalline regions, preferred for X and gamma radiation detectors, are highlighted in white and represent a small portion of the sample. The majority of the sample consists of blocks, which are less desirable in the aforementioned application because crystal structure defects serve as scattering or recombination centers for charge carriers.

Omega scans for selected points X, Y: (−7, −6) and (6, −2) are presented in Fig. 6c and Fig. 6d, respectively. In Fig. 6c, four maxima can be observed, signifying a block-like or sub-grain structure of the sample. The FWHM value taken from the extreme maxima is 300 arcsec. Meanwhile, in Fig. 6d, a single, very narrow peak with an FWHM of 15 arcsec is seen, indicating a monocrystalline structure of the sample at that particular location. In this paper, we observed the same issues that our (Cd,Mn)(Te,Se) crystals had previously encountered [37]. Although the sample appeared to be monocrystalline during visual observation of the Durose-etched surface, revealing no grains and twins, it is, in fact, composed of misoriented blocks with a significant mosaic component (areas with a surface on the order of square millimeters).

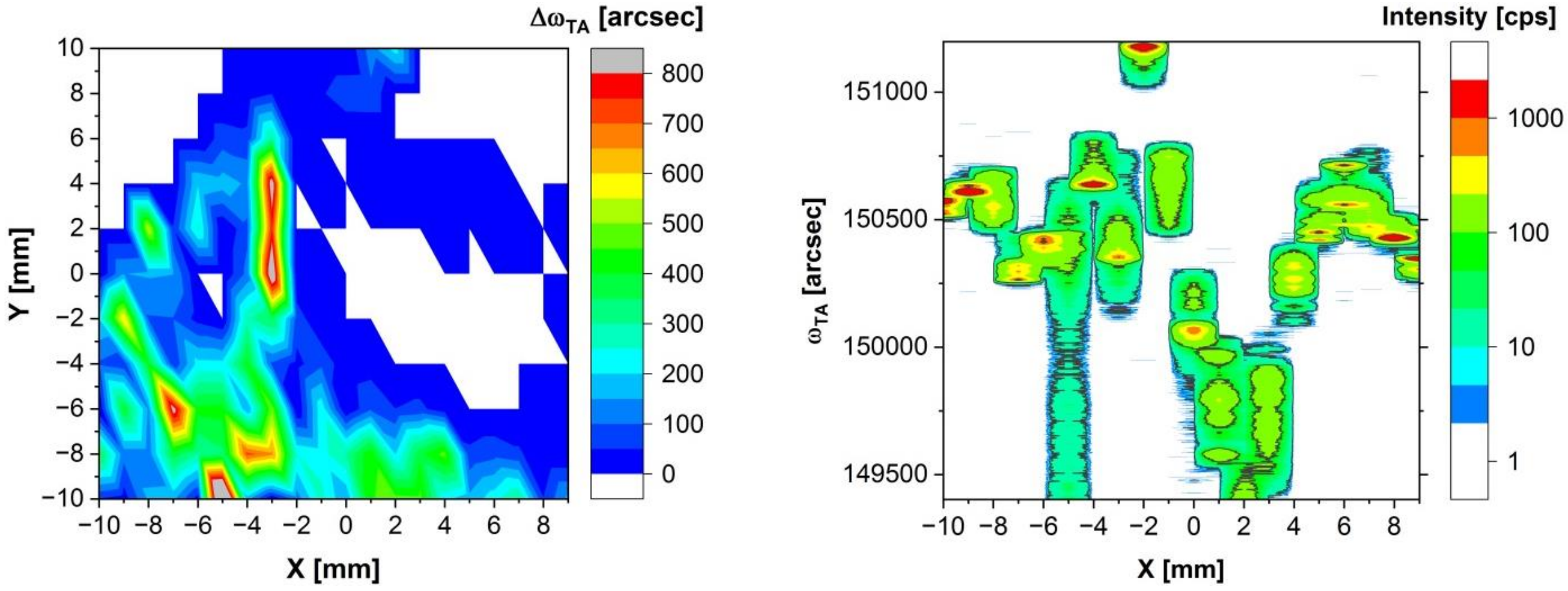

(a) (b)

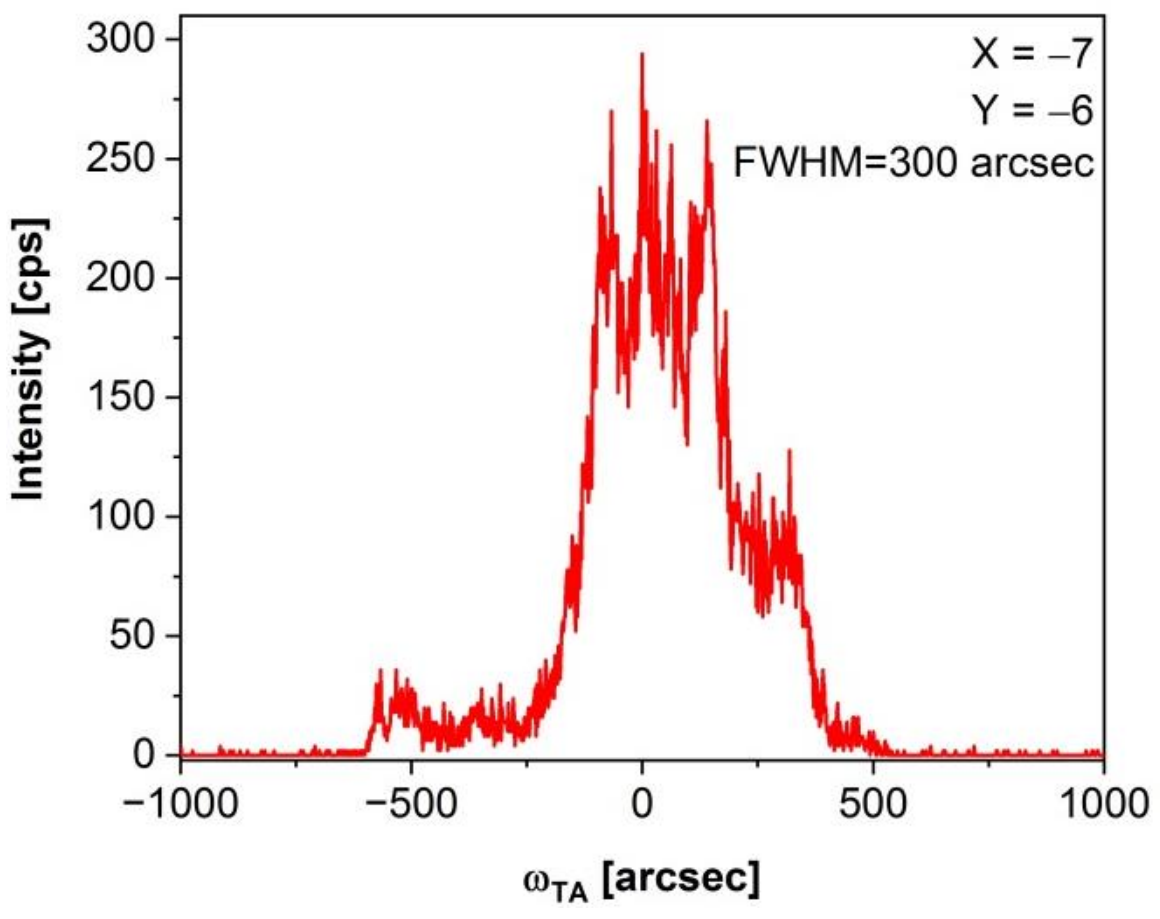

(c)

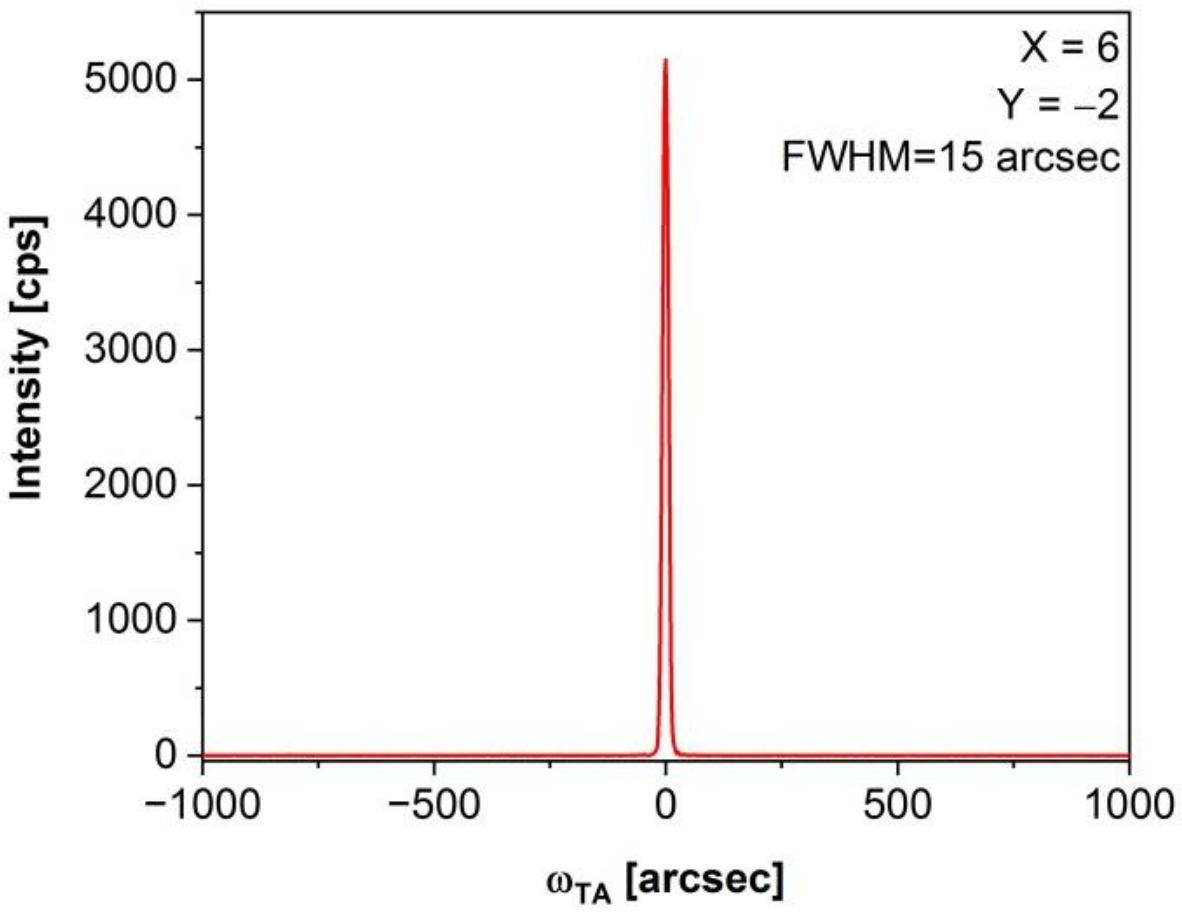

(d)

Fig. 6. $Cd_{0.93}Mn_{0.07}Te_{0.98}Se_{0.02}$ results. (a) Delta omega map of $Cd_{0.93}Mn_{0.07}Te_{0.98}Se_{0.02}$ in the triple axis mode. (b) The map of the intensity of omega values, $\omega_{TA}$, obtained from 20 omega scans conducted along the Y = −10 line, with a 1 mm X step. (c) Omega scan for measurement point X, Y: (−7, −6). Four distinct maxima are visible, indicating a block-like or sub-grain structure of the sample. (d) Omega scan for measurement point X, Y: (6, −2). A narrow peak with an FWHM of 15 arcsec indicates a monocrystalline structure of the sample at that particular location. It is worth noting that the scale of the X axis for Fig. 6c and d is identical. In the (Cd,Mn)(Te,Se) sample, there are areas with block-like structures (Fig. 6c) as well as perfectly monocrystalline regions (Fig. 6d).

According to Darwin's model [62], a monocrystal is composed of a mosaic (blocks) with sizes ranging from 10 nm to 1 μm, slightly misoriented with respect to each other. The angle of misorientation between blocks typically ranges from a few arcseconds to a few minutes, in exceptional cases a few degrees. These small-angle boundaries are formed by a set of dislocations.

Micro-mosaic is present in practically every crystal. However, delta omega maps of our (Cd,Mn)Te (Fig. 5a) and (Cd,Mn)(Te,Se) (Fig. 6a) samples indicate a significantly higher contribution of block-like structure in the second one. Furthermore, in the (Cd,Mn)(Te,Se) crystal, the maximum misorientation between the blocks, $\Delta\omega_{TA}$, is 10 times larger than in the (Cd,Mn)Te crystal. Therefore, the results of omega scans suggest a more perfect crystalline structure in the crystal without selenium alloying, i.e., (Cd,Mn)Te.

### 3.2.4 FWHM of Omega Scans

Now, let us consider the (Cd,Mn)Te crystal once again. In Fig. 7a, a map of the FWHM obtained from the omega scans at consecutive points of the sample is presented. This map was obtained in triple-axis mode. Several (five) points with worse (higher) FWHM values are located in the lower-right corner, which corresponds well to Fig. 5a, where the presence of two blocks was recorded in that area. Apart from these exceptions, in the monocrystalline region of the sample, the FWHM of the omega scan is consistently better than ~50 arcsec.

In Fig. 7b, a selected omega scan for the measurement point $X = -1$, $Y = -6$, chosen from the map in Fig. 7a, has been presented. When measured in double-axis (DA) mode, meaning without the use of an analyzer, the FWHM of this rocking curve is 38 arcsec. However, when an analyzer is employed, i.e., in triple-axis mode, the FWHM is reduced to 20 arcsec.

An FWHM map is not presented for the (Cd,Mn)(Te,Se) crystal because the omega scans resulted in curves with multiple maxima (indicating the presence of blocks/grains in the sample). Therefore, it is challenging to arbitrarily determine which FWHM value should be included in the map.

For comparison, high-resolution rocking curve measurements of THM-grown (Cd,Zn)(Te,Se) crystals resulted in an FWHM value of 30.8 arcsec, and no mosaic structure was observed [63]. This outcome can be attributed to the THM method's slower growth rate compared to the Bridgman method, mentioned in Section 1, which leads to fewer structural defects and, consequently, the achievement of a low FWHM value. On the other hand, when examining rocking curve studies of Bridgman-grown crystals, a broad spectrum of reported FWHM values for omega scans in the case of (Cd,Zn)Te exists, ranging from 8 to over 400 arcsec [64–66]. For as-grown (Cd,Mn)Te crystals, previous research has reported FWHM values of 68 arcsec [67] or 72 arcsec [68]. In contrast, our (Cd,Mn)Te crystal demonstrates superior performance, featuring an FWHM value for the omega scan that is almost two times smaller (in double-axis mode).

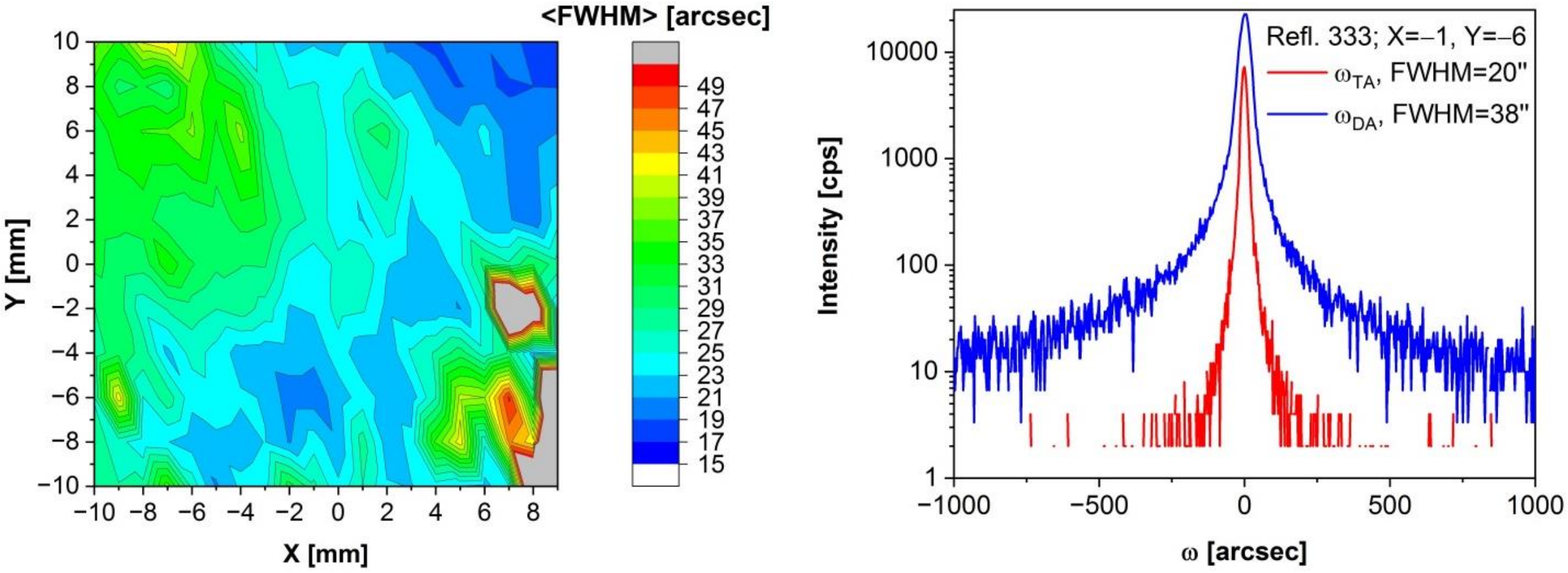

(a) (b)

Fig. 7. (a) Map of FWHM for $Cd_{0.95}Mn_{0.05}Te$ acquired through omega scans conducted using triple-axis mode. (b) FWHM comparison at X = −1, Y = −6: 38 arcsec (double-axis, DA), 20 arcsec (triple-axis, TA).

Our X-ray examinations suggest a better crystal structure in (Cd,Mn)Te crystals compared to (Cd,Mn)(Te,Se) crystals. The distribution of lattice constant in both samples was very good, exhibiting minimal changes at the ppm level. However, omega scans revealed a significant presence of block/grain-like structures in (Cd,Mn)(Te,Se) crystals, much higher than in (Cd,Mn)Te crystals, and displayed a higher degree of misorientation. Both X-ray studies and EPD (Etch Pit Density) measurements suggest that (Cd,Mn)Te crystals are more suitable for X-ray and gamma detectors compared to crystals with selenium addition.

### 3.3 Impact of grain boundaries and twins

The influence of grain boundaries and twins on the FWHM of the omega curve measurement was investigated in (Cd,Mn)Te crystals. Specifically for this purpose, a (Cd,Mn)Te plate with both grain boundary and twin was examined. Studying grain boundaries and twins is crucial in CdTe-based materials for X-ray and gamma-ray detectors because understanding the structure of grain boundaries and twins can lead to improvements in the detector manufacturing process and the quality of X-ray and gamma-ray radiation detection, including the energy resolution of the detector.

Fig. 8a presents a compilation of several infrared (IR) images, each focused at different depths within the sample, in order to illustrate the width of the grain boundary. This grain boundary has a plane that is inclined relative to the imaging plane. The dark, spherical objects visible in Fig. 8a correspond to tellurium inclusions situated within the grain boundary region, as investigated by us in Refs. [69,70]. In this specific area, the width of the grain boundary measures 900 μm. Fig. 8b shows changes in the FWHM values of the omega curves, recorded at intervals of 0.2 mm along the sample, as it transitions from one grain to another. Notably, the region where the FWHM undergoes a significant shift of approximately 60-70 arcsec spans a width of 1.6 mm. It is worth noting that the X-ray scan was conducted at a slightly different location compared to the IR image, which accounts for the variance in the grain boundary width values between the X-ray data and the IR image.

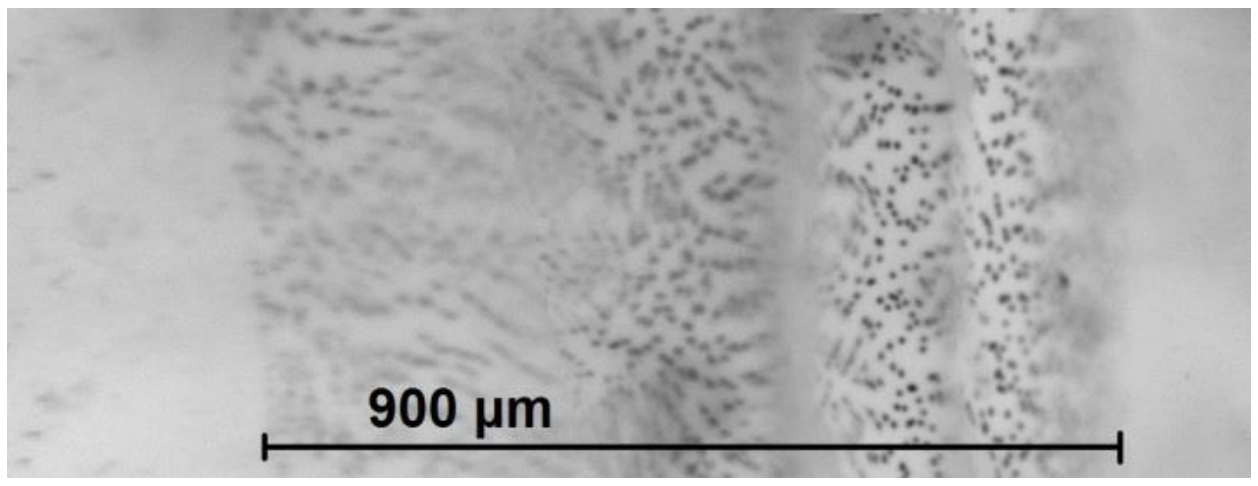

(a)

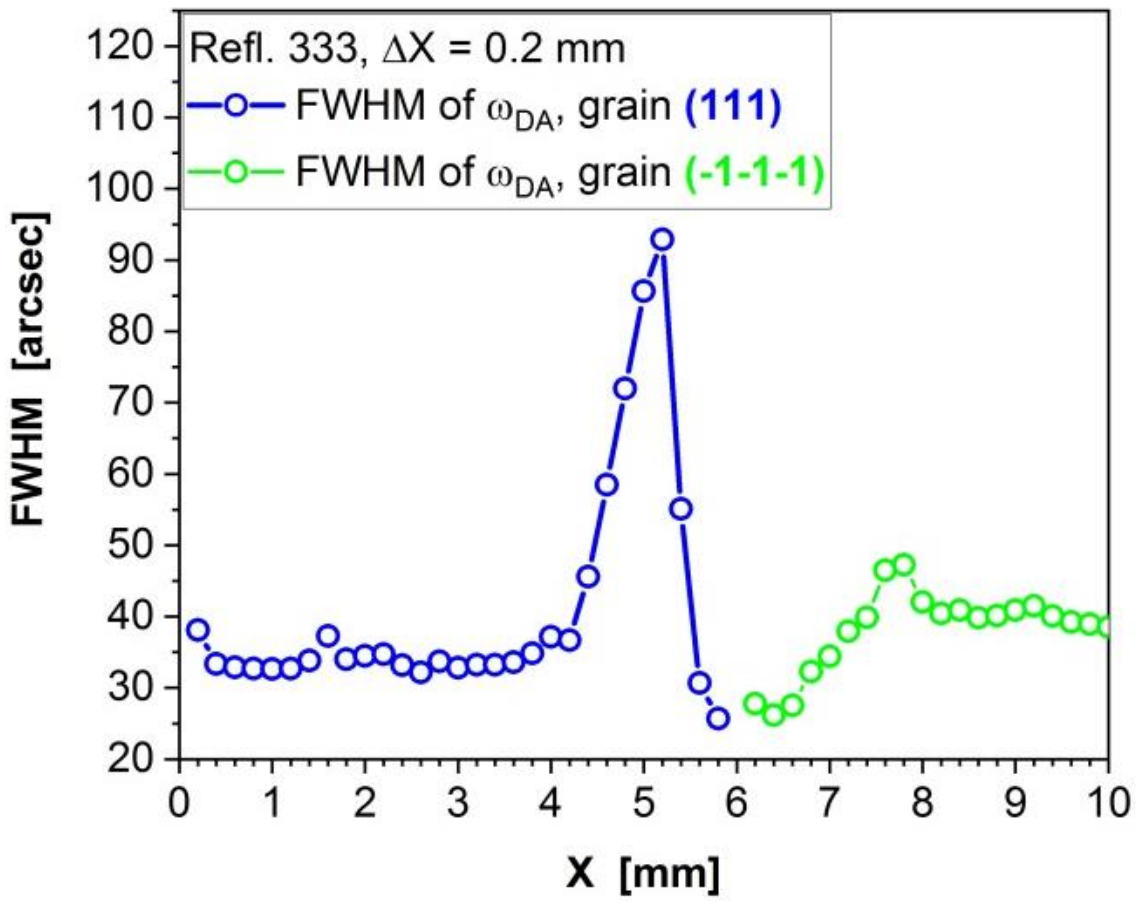

(b)

Fig. 8. Harmful influence of grain boundary in (Cd,Mn)Te crystals. (a) Fused infrared images, focused at various depths of the sample, revealing the grain boundary width in this specimen. (b) FWHM of omega curve as a function of the measurement point along the sample. Due to the steep inclination of adjacent grains, the sample was scanned along a line from both sides of the grain boundary – from the left side of the grain boundary, corresponding to the blue points on the curve, as well as from the right side of the grain boundary, corresponding to the green points on the curve.

Fig. 9a displays an IR image of a twin in (Cd,Mn)Te. This twin is decorated with tellurium inclusions, visible as dark objects arranged vertically on the left side of the image. The width of the twin measures approximately 70 μm and is one order of magnitude smaller than the width of the grain boundary shown in Fig. 8a.

Fig. 9b presents the FWHM values of the omega curves obtained during the scanning of a (111)-oriented sample with a 0.2 mm step, transitioning from one part of the twin to the other. In this case, the FWHM values are consistently below 12 arcsec, which is close to the limit of our diffractometer's resolution in TA mode (8 arcsec). Importantly, no significant changes in FWHM values are observed within the standard error. This is because the two crystal parts separated by the twin exhibit high crystallographic quality in the measurement area, and they are rotated relative to each other around the normal to the (111) plane. As anticipated, the presence of the twin in the crystal does not seem to impact the FWHM, especially when the boundary contains few defects.

Both techniques, infrared imaging and X-ray diffraction studies, underscore the detrimental impact of grain boundaries and the negligible effect of twins on the discussed properties of (Cd,Mn)Te crystals.

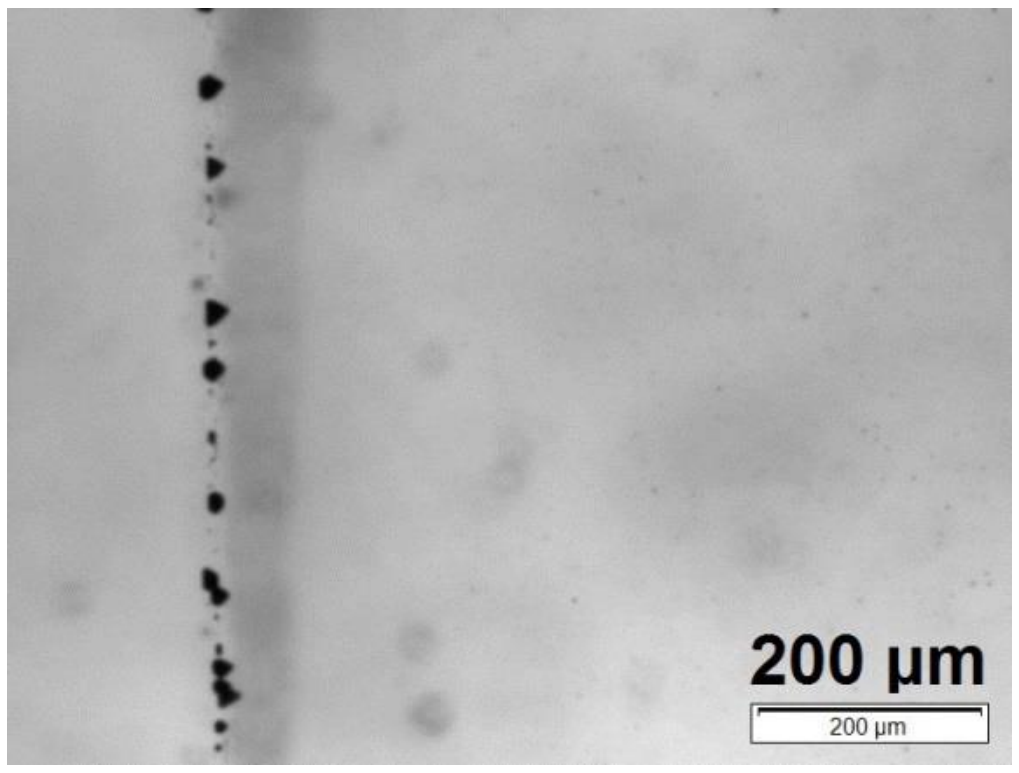

(a)

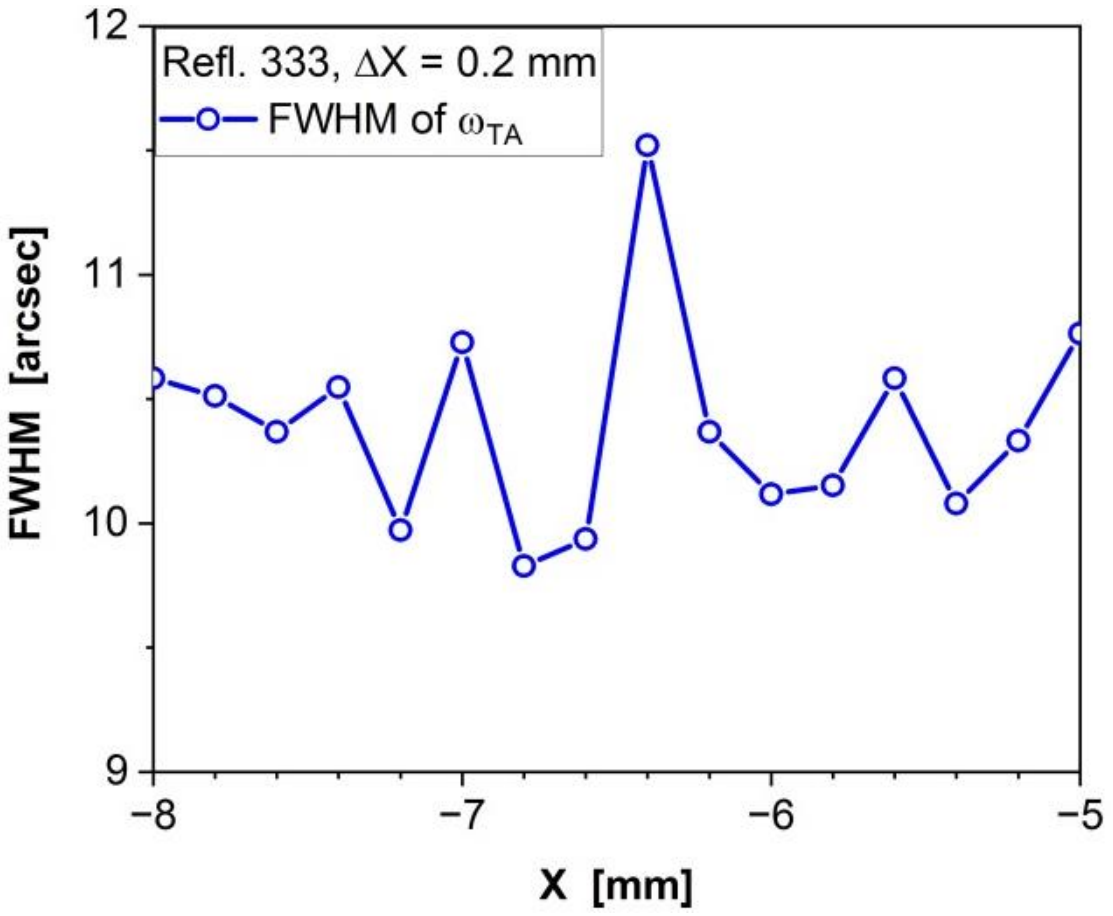

(b)

Fig. 9. Negligible effect of twin on FWHM of omega curve in (Cd,Mn)Te crystal. (a) Infrared image of a twin decorated with tellurium inclusions, visible as dark objects aligning in a row on the left side of the image. (b) FWHM of omega curve as a function of the measurement point along the sample.

### 3.4 Photoluminescence of as-grown and annealed crystals

Fig. 10a and 10b depict low-temperature photoluminescence (PL) spectra of (Cd,Mn)Te and (Cd,Mn)(Te,Se) samples, respectively. These spectra exhibit common features with the PL spectra of (Cd,Zn)Te. In both the (Cd,Mn)Te and (Cd,Mn)(Te,Se) samples we investigated, we can identify excitonic luminescence, donor-acceptor transitions, and defect-related bands, similar to what is observed in (Cd,Zn)Te [39].

In both of the as-grown materials, there are excitonic transitions, including $D^0X$ (exciton bound to a neutral donor) and $A^0X$ (exciton bound to a neutral acceptor), as well as two donor-acceptor pair transitions (DAP). In some instances, these transitions are accompanied by their phonon replicas, with energies approximately 20 meV lower. Specifically, shallow (s) and deep (d) DAP transitions are located about 70 meV and 200 meV below the exciton lines, respectively.

Bridgman-grown (Cd,Mn)Te and (Cd,Mn)(Te,Se) crystals naturally exhibit a high concentration of Cd vacancies, which act as acceptors. This is a consequence of the insufficient Cd content at high temperature during crystal growth (~1100 °C), caused by the high partial pressure of Cd. To reduce the concentration of cadmium vacancies, we applied annealing to both crystals, with and without selenium, in a cadmium-rich environment.

Consequently, in (Cd,Mn)Te crystals, the intensity of the $A^0X$ and $DAP^s$ PL lines was reduced, and emission from a $DAP^d$ transition was eliminated, as demonstrated in Fig. 10a. Thus, it can be inferred that the concentration of acceptors is lower in the Cd-annealed sample compared to the as-grown one. Previous studies have shown that annealing in Cd vapors effectively eliminates the PL peak associated with the $DAP^s$ transition in our (Cd,Mn)Te samples, as well (Figs. 9 and 10 in [71]).

Annealing of (Cd,Mn)(Te,Se) crystals in cadmium vapors also led to a reduction in the PL intensity of $DAP^s$ and $DAP^d$ lines, consequently decreasing the concentration of cadmium vacancy acceptors, which is presented in Fig. 10b. However, this effect is notably less pronounced compared to (Cd,Mn)Te crystals. On the other hand, our prior research demonstrated that subjecting (Cd,Mn)(Te,Se) samples to double annealing in cadmium vapors had a negligible impact on the intensity of the $DAP^s$ and $DAP^d$ PL lines (as shown in Fig. 8 in [37]).

Conversely, in the (Cd,Mn)(Te,Se) crystal the annealing process in selenium vapors primarily affected the concentration of donors (Fig. 10b). Evaluating the change in PL intensity of $DAP^s$ and $DAP^d$ lines after annealing in selenium vapors presents a challenge due to the disappearance of the reference line, $D^0X$. It is likely that the number of selenium vacancies, which should, in principle, act as deep donors (similar to tellurium vacancies [37,72,73]), has been reduced. Furthermore, in (Cd,Mn)(Te,Se), the $DAP^d$ PL line is observed at higher temperatures, extending up to 120 K, compared to (Cd,Mn)Te, where it is observed up to 100 K. In both crystals, the $DAP^s$ emission disappears at 60 K.

When comparing the PL spectra of both (Cd,Mn)Te and (Cd,Mn)(Te,Se) crystals, it can be observed that in (Cd,Mn)Te crystals, changes in the intensity of PL lines associated with acceptors ($A^0X$, DAP) and, consequently, changes in the concentration of acceptors (cadmium vacancies) after annealing in cadmium vapors are more noticeable than in Cd-annealed (Cd,Mn)(Te,Se) crystals. In (Cd,Mn)(Te,Se) crystals, there may exist complexes containing selenium vacancies, which create deep energy levels, i.e., deep traps for charge carriers. This could potentially explain the difficulties encountered in eliminating the $DAP^s$ and $DAP^d$ PL lines in the spectra of crystals containing selenium.

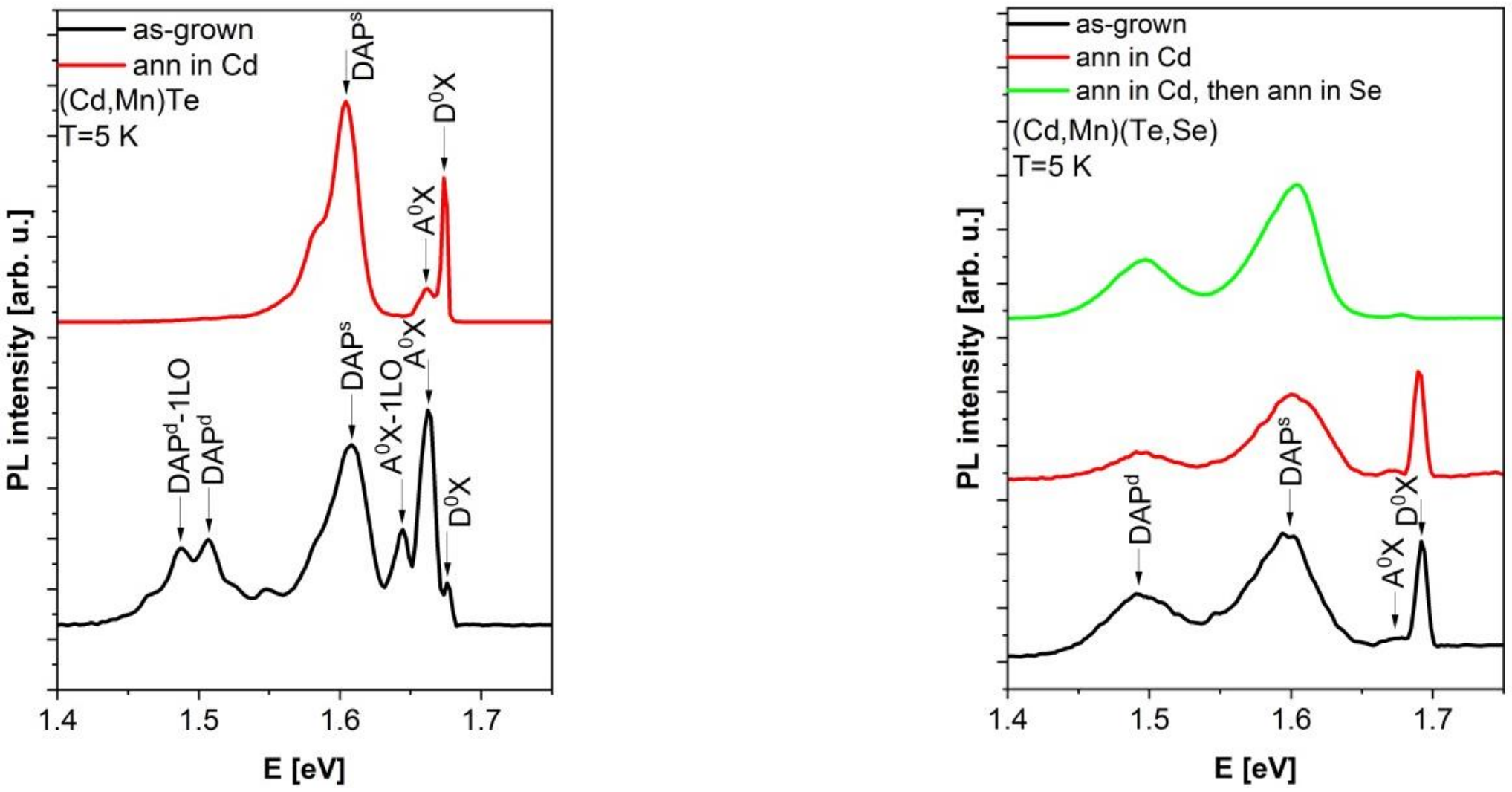

(a) (b)

Fig. 10. Photoluminescence spectra at the temperature of 5 K. (a) $Cd_{0.95}Mn_{0.05}Te$ sample. Annealing in Cd vapors eliminated the $DAP^d$ luminescence. (b) $Cd_{0.95}Mn_{0.05}Te_{0.98}Se_{0.02}$ sample. Annealing in Cd or Se vapors did not eliminate the $DAP^s$ and $DAP^d$ luminescence.

### 3.5 Detector response

Finally, a comparison was made between the detector response of two materials, (Cd,Mn)Te and (Cd,Mn)(Te,Se). The performance of the detectors at room temperature was assessed using a Co-57 point source. An example image of a (Cd,Mn)Te pixelated detector, which was prepared in our laboratory, is shown in Fig. 11a.

Our as-grown (Cd,Mn)Te detector is capable of distinguishing 122 keV gamma-rays from Co-57 with an energy resolution ranging from 8% to 17%. The spectroscopic performance of a selected (Cd,Mn)Te detector pixel, featuring an FWHM of 14%, is presented in Fig. 11b.

Conversely, our (Cd,Mn)(Te,Se) detector only detects X-rays from Co-57 at 7 keV with an energy resolution of approximately 45%, along with a minor trace of gamma-rays at 14.4 keV. The reduced performance of the (Cd,Mn)(Te,Se) detector may be associated with the presence of a deep trap contributing to the luminescence of DAP[d], and a substantial presence of blocks in the crystal structure, although further investigation is needed to confirm this hypothesis.

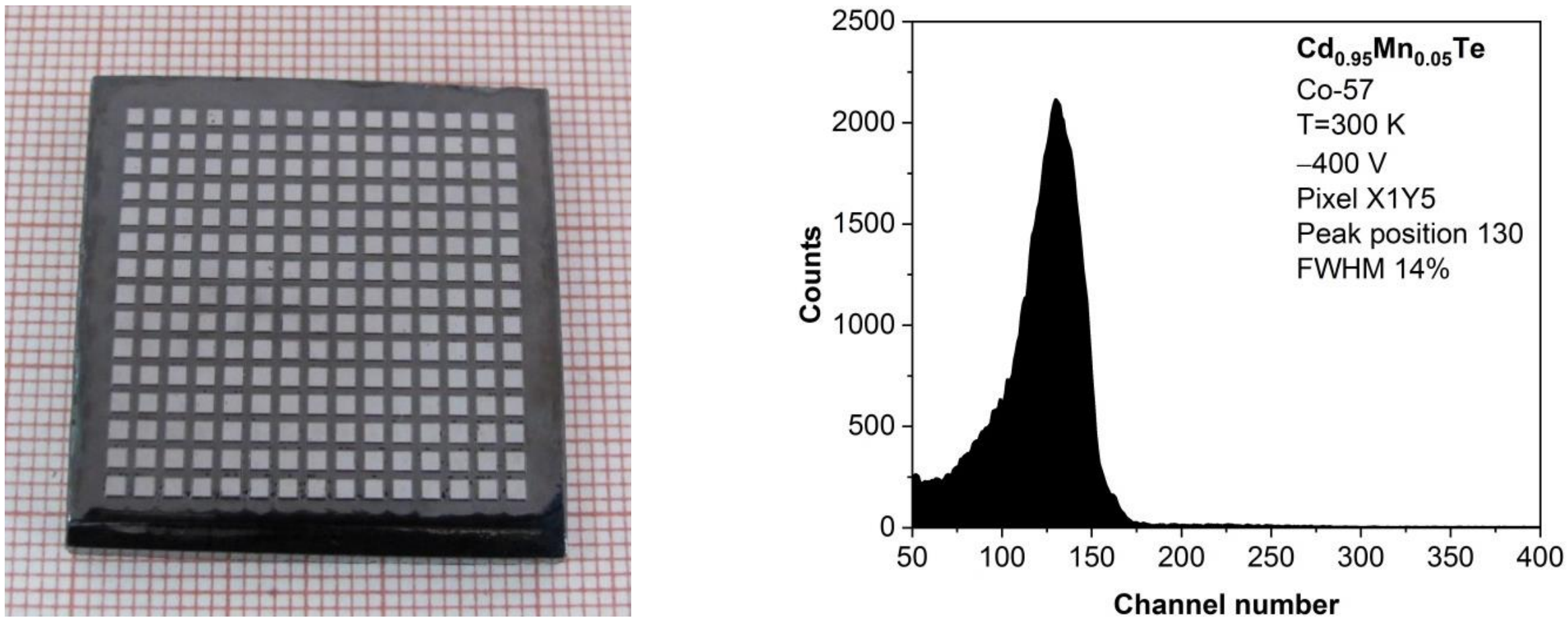

(a) (b)

Fig. 11. As-grown $Cd_{0.95}Mn_{0.05}Te$ pixelated detector. (a) A photograph of the detector with dimensions of 32×32 mm² and 225 pixels. (b) Spectroscopic performance of a from a selected pixel made at room temperature using a Co-57 source. The cathode was biased with −400 V. The peak in the spectrum is related to 122 keV.

## 4. Conclusions

We conducted a comparative analysis of two CdTe-based compounds, (Cd,Mn)Te and (Cd,Mn)(Te,Se), both grown using the Bridgman method, focusing on their crystal structure, hardness, luminescence properties, and their effectiveness as X-ray and gamma-ray detectors.

X-ray examinations of visually identified monocrystalline samples revealed very uniform lattice constants in both crystals, with minimal variations at the ppm level. However, omega curve measurements unveiled a significant presence of block-like structures within (Cd,Mn)(Te,Se) crystals, resulting in delta omega values, corresponding to the maximum misorientation between blocks, on the order of 100 arcsec (with a peak at 800 arcsec). In contrast, (Cd,Mn)Te crystals exhibited nearly perfect monocrystalline structures, with block-like features observed in only 2% of the 18×20 mm$^2$ area. Additionally, the misorientation angles between blocks in (Cd,Mn)Te were approximately ten times smaller than those observed in the selenium-containing crystals. Etching the crystals with Inoue solution further emphasized this contrast, displaying one order of magnitude fewer etch pits in (Cd,Mn)Te compared to (Cd,Mn)(Te,Se). The study also highlighted the detrimental influence of grain boundaries and the negligible impact of twins on the crystal structure quality of our samples.

We find that (Cd,Mn)Te shows greater promise as a material for X-ray and gamma-ray detectors. It exhibits the ability to distinguish 122 keV gamma-rays from a Co-57 source with an energy resolution of 8-17%. Conversely, our (Cd,Mn)(Te,Se) detectors exhibited poor responses to X- and gamma-rays, potentially due to the presence of a deep trap involved in DAP[d] luminescence, which cannot be eliminated through annealing in Cd vapors, unlike in the case of (Cd,Mn)Te. Additionally, the significant contribution of block-like structures in selenium-containing crystal samples, accompanied by notably larger misorientation angles between these blocks compared to (Cd,Mn)Te, may contribute to the bad performance.

**Funding sources**

This work was supported by the Polish National Centre for Research and Development, grant number TECHMATSTRATEG1/346720/8/NCBR/2017, and by the Foundation for Polish Science through the IRA Program co-financed by European Union within SG OP, grant number MAB/2017/1.

**CRediT authorship contribution statement**

**Aneta Masłowska:** Conceptualization, Methodology, Investigation, Resources, Writing – Original Draft, Visualization; **Dominika M. Kochanowska:** Methodology, Investigation, Resources, Writing – Review & Editing; **Adrian Sulich:** Formal analysis, Writing - Review & Editing, Visualization; **Jaroslaw Z. Domagala:** Investigation, Methodology, Writing – Review & Editing, Visualization; **Marcin Dopierała:** Validation, Resources; **Michał Kochański:** Resources; **Michał Szot:** Investigation, Validation; **Witold Chromiński:** Investigation; **Andrzej Mycielski:** Conceptualization, Resources, Writing – Review & Editing, Supervision, Funding acquisition

## Acknowledgment

The authors would like to thank Janusz Gdański, Stanisław Jabłoński, Adam Marciniak, Paweł Skupiński, and Marek Zubrzycki for their technical support during experiments.

## References

[1] A. Hossain, A.E. Bolotnikov, G.S. Camarda, Y. Cui, D. Jones, J. Hall, K.H. Kim, J. Mwathi, X. Tong, G. Yang, R.B. James, Novel approach to surface processing for improving the efficiency of CdZnTe detectors, J. Electron. Mater. 43 (2014) 2771–2777. https://doi.org/10.1007/s11664-013-2698-5.

[2] L. Abbene, G. Gerardi, A.A. Turturici, G. Raso, G. Benassi, M. Bettelli, N. Zambelli, A. Zappettini, F. Principato, X-ray response of CdZnTe detectors grown by the vertical Bridgman technique: Energy, temperature and high flux effects, Nucl. Instruments Methods Phys. Res. Sect. A Accel. Spectrometers, Detect. Assoc. Equip. 835 (2016) 1–12. https://doi.org/10.1016/j.nima.2016.08.029.

[3] M.C. Veale, P. Booker, S. Cross, M.D. Hart, L. Jowitt, J. Lipp, A. Schneider, P. Seller, R.M. Wheater, M.D. Wilson, C.C.T. Hansson, K. Iniewski, P. Marthandam, G. Prekas, Characterization of the uniformity of high-flux CdZnTe material, Sensors (Switzerland). 20 (2020) 2747. https://doi.org/10.3390/s20102747.

[4] M.D. Alam, S.S. Nasim, S. Hasan, Recent progress in CdZnTe based room temperature detectors for nuclear radiation monitoring, Prog. Nucl. Energy. 140 (2021) 103918. https://doi.org/https://doi.org/10.1016/j.pnucene.2021.103918.

[5] M. Rejhon, J. Franc, V. Dědič, J. Pekárek, U.N. Roy, R. Grill, R.B. James, Influence of deep levels on the electrical transport properties of CdZnTeSe detectors, J. Appl. Phys. 124 (2018) 235702. https://doi.org/10.1063/1.5063850.

[6] U.N. Roy, G.S. Camarda, Y. Cui, R.B. James, High-resolution virtual Frisch grid gamma-ray detectors based on as-grown CdZnTeSe with reduced defects, Appl. Phys. Lett. 114 (2019) 232107. https://doi.org/10.1063/1.5109119.

[7] U.N. Roy, G.S. Camarda, Y. Cui, R.B. James, Optimization of selenium in CdZnTeSe quaternary compound for radiation detector applications, Appl. Phys. Lett. 118 (2021) 152101. https://doi.org/10.1063/5.0048875.

[8] U.N. Roy, G.S. Camarda, Y. Cui, G. Yang, R.B. James, Impact of selenium addition to the cadmium-zinc-telluride matrix for producing high energy resolution X-and gamma-ray detectors, Sci. Rep. 11 (2021) 10338. https://doi.org/10.1038/s41598-021-89795-z.

[9] M. Rejhon, V. Dedic, R. Grill, J. Franc, U.N. Roy, R.B. James, Low-temperature annealing of CdZnTeSe under bias, Sensors. 22 (2022) 171. https://doi.org/10.3390/s22010171.

[10] U.N. Roy, G.S. Camarda, Y. Cui, R. Gul, A. Hossain, G. Yang, J. Zazvorka, V. Dedic, J. Franc, R.B. James, Role of selenium addition to CdZnTe matrix for room-temperature radiation detector applications, Sci. Rep. 9 (2019) 1620. https://doi.org/10.1038/s41598-018-38188-w.

[11] P. Moravec, J. Franc, V. Dedic, P. Minarik, H. Elhadidy, V. Sima, R. Grill, U. Roy, Microhardness study of CdZnTeSe crystals for X-ray and gamma ray radiation detectors, 2019 IEEE Nucl. Sci. Symp. Med. Imaging Conf. NSS/MIC 2019. (2019) 2–5. https://doi.org/10.1109/NSS/MIC42101.2019.9059851.

[12] M. Rejhon, V. Dedic, L. Beran, U.N. Roy, J. Franc, R.B. James, V. Dědič, L. Beran, U.N. Roy, J. Franc, R.B. James, Investigation of Deep Levels in CdZnTeSe Crystal and Their Effect on the Internal Electric Field of CdZnTeSe Gamma-Ray Detector, IEEE Trans. Nucl. Sci. 66 (2019) 1952–1958. https://doi.org/10.1109/TNS.2019.2925311.

[13] S.U. Egarievwe, U.N. Roy, C.A. Goree, B.A. Harrison, J. Jones, R.B. James, Ammonium fluoride passivation of CdZnTeSe sensors for applications in nuclear detection and medical imaging, Sensors. 19 (2019) 3271.

[14] S.K. Chaudhuri, M. Sajjad, J.W. Kleppinger, K.C. Mandal, Charge transport properties in CdZnTeSe semiconductor room-temperature γ-ray detectors, J. Appl. Phys. 127 (2020) 245706.

[15] M. Sajjad, S.K. Chaudhuri, J.W. Kleppinger, O. Karadavut, K.C. Mandal, Investigation on Cd0.9Zn0.1Te1-ySey single crystals grown by vertical Bridgman technique for high-energy

gamma radiation detectors, Hard X-Ray, Gamma-Ray, Neutron Detect. Phys. XXII. 11494 (2020) 114941F. https://doi.org/10.1117/12.2570592.

[16] U.N. Roy, G.S. Camarda, Y. Cui, R.B. James, Advances in CdZnTeSe for Radiation Detector Applications, Radiation. 1 (2021) 123–130. https://doi.org/10.3390/radiation1020011.

[17] J. Pipek, M. Betušiak, E. Belas, R. Grill, P. Praus, A. Musiienko, J. Pekarek, U.N. Roy, R.B. James, Charge Transport and Space-Charge Formation in Cd1-xZnxTe1-ySey Radiation Detectors, Phys. Rev. Appl. 15 (2021) 054058. https://doi.org/10.1103/PhysRevApplied.15.054058.

[18] S.U. Egarievwe, U.N. Roy, E.O. Agbalagba, K.L. Dunning, O.K. Okobiah, M.B. Israel, M.L. Drabo, R.B. James, Characterization of CdMnTe Planar Nuclear Detectors Grown by Vertical Bridgman Technique, in: 2019 IEEE Nucl. Sci. Symp. Med. Imaging Conf., IEEE, 2019: pp. 1–3.

[19] S.U. Egarievwe, R.B. James, M.B. Israel, A.D. Banks, M.L. Drabo, K.L. Dunning, V.J. Cook, F.D. Johnson, S.M. Palmer, U.N. Roy, R.B. James, Design and fabrication of a CdMnTe nuclear radiation detection system, in: 2019 SoutheastCon, IEEE, 2019: pp. 1–4. https://doi.org/10.1109/SoutheastCon42311.2019.9020612.

[20] L. Luan, H. Lv, L. Gao, Y. He, D. Zheng, Preparation and properties of hemispherical CdMnTe nuclear radiation detectors, Nucl. Instruments Methods Phys. Res. B. 471 (2020) 42–47. https://doi.org/10.1016/j.nimb.2020.03.018.

[21] S.U. Egarievwe, M.B. Israel, A. Davis, M. McGuffie, K. Hartage, M.A. Alim, U.N. Roy, R.B. James, X-ray Photoelectron Spectroscopy of CdZnTe and CdMnTe Materials for Nuclear Detectors, 2020 IEEE Nucl. Sci. Symp. Med. Imaging Conf. NSS/MIC 2020. (2020) 1–3. https://doi.org/10.1109/NSS/MIC42677.2020.9508043.

[22] P. Yu, P. Gao, T. Shao, W. Liu, B. Jiang, C. Liu, Z. Ma, J. Zheng, Correlation between Te inclusions and the opto-electrical properties of CdMnTe and CdMgTe single crystals, J. Cryst. Growth. 571 (2021) 126259. https://doi.org/10.1016/j.jcrysgro.2021.126259.

[23] P. Yu, T. Shao, Z. Ma, P. Gao, B. Jing, W. Liu, C. Liu, Y. Chen, Y. Liu, Z. Fang, L. Luan, Influence of hydrogen treatment on electrical properties of detector-grade CdMnTe:In crystals,

IEEE Trans. Nucl. Sci. 68 (2021) 458–462. https://doi.org/10.1109/TNS.2021.3067726.

[24] J. Byun, J. Seo, B. Park, Growth and characterization of detector-grade CdMnTeSe, Nucl. Eng. Technol. 54 (2022) 4215–4219. https://doi.org/10.1016/j.net.2022.06.007.

[25] Y. Kim, J. Ko, J. Byun, J. Seo, B. Park, Passivation effect on Cd0.95Mn0.05Te0.98Se0.02 radiation detection performance, Appl. Radiat. Isot. 200 (2023) 110914. https://doi.org/10.1016/j.apradiso.2023.110914.

[26] A. Hossain, V. Yakimovich, A.E. Bolotnikov, K. Bolton, G.S. Camarda, Y. Cui, J. Franc, R. Gul, K.H. Kim, H. Pittman, G. Yang, R. Herpst, R.B. James, Development of Cadmium Magnesium Telluride (Cd1-xMg xTe) for room temperature X- and gamma-ray detectors, J. Cryst. Growth. 379 (2013) 34–40. https://doi.org/10.1016/j.jcrysgro.2012.11.044.

[27] S.B. Trivedi, S.W. Kutcher, W. Palosz, M. Berding, A. Burger, W. Palsoz, M. Berding, A. Burger, Next Generation Semiconductor-Based Radiation Detectors Using Cadmium Magnesium Telluride, in: Brimrose Technology Corporation, Sparks Glencoe, MD (United States), 2014. https://doi.org/https://doi.org/10.2172/1165052.

[28] A. Mycielski, D.M. Kochanowska, M. Witkowska-Baran, A. Wardak, M. Szot, J. Domagała, B.S. Witkowski, R. Jakieła, L. Kowalczyk, B. Witkowska, Investigation of Cd1−xMgxTe as possible materials for X and gamma ray detectors, J. Cryst. Growth. 491 (2018) 73–76. https://doi.org/10.1016/j.jcrysgro.2018.03.035.

[29] P. Yu, B. Jiang, Y. Chen, J. Zheng, L. Luan, Study on In-Doped CdMgTe Crystals Grown by a Modified Vertical Bridgman Method Using the ACRT Technique, Materials (Basel). 12 (2019) 4236. https://doi.org/10.3390/ma1224236.

[30] P. Yu, B. Jiang, Z. Han, S. Zhao, P. Gao, T. Shao, W. Liu, X. Gu, Y. Wang, Characterization of physical and optical properties of a new radiation detection material CdMgTe, Opt. Mater. (Amst). 131 (2022) 112656. https://doi.org/10.1016/j.optmat.2022.112656.

[31] P. Yu, P. Gao, B. Jiang, Z. Han, S. Zhao, W. Liu, X. Sun, L. Luan, T. Rao, Effects of electrode fabrication on electrical properties of CdMgTe room temperature radiation detectors, Mater. Sci.

Semicond. Process. 153 (2023) 107178. https://doi.org/10.1016/j.mssp.2022.107178.

[32] G. Camarda, G. Yang, A.E. Bolotnikov, Y. Cui, A. Hossain, K.H. Kim, U. Roy, R.B. James, Characterization of Detector-Grade CdTeSe Crystals, Brookhaven National Lab.(BNL), Upton, NY (United States), 2013.

[33] U.N. Roy, A.E. Bolotnikov, G.S. Camarda, Y. Cui, A. Hossain, K. Lee, M. Marshall, G. Yang, R.B. James, Growth of CdTexSe1− x from a Te-rich solution for applications in radiation detection, J. Cryst. Growth. 386 (2014) 43–46.

[34] U.N. Roy, A.E. Bolotnikov, G.S. Camarda, Y. Cui, A. Hossain, K. Lee, G. Yang, R.B. James, Evaluation of CdTexSe1− x crystals grown from a Te-rich solution, J. Cryst. Growth. 389 (2014) 99–102.

[35] U.N. Roy, A.E. Bolotnikov, G.S. Camarda, Y. Cui, A. Hossain, K. Lee, W. Lee, R. Tappero, G. Yang, R. Gul, R.B. James, High compositional homogeneity of CdTexSe1− x crystals grown by the Bridgman method, APL Mater. 3 (2015) 26102.

[36] R. Gul, U.N. Roy, S.U. Egarievwe, A.E. Bolotnikov, G.S. Camarda, Y. Cui, A. Hossain, G. Yang, R.B. James, Point defects: Their influence on electron trapping, resistivity, and electron mobility-lifetime product in CdTexSe1− x detectors, J. Appl. Phys. 119 (2016) 25702.

[37] A. Mycielski, D. Kochanowska, M. Witkowska-Baran, A. Wardak, M. Szot, J.Z. Domagała, R. Jakieła, L. Kowalczyk, B. Witkowska, Semiconductor crystals based on CdTe with Se–Some structural and optical properties, J. Cryst. Growth. 498 (2018) 405–410.

[38] U.N. Roy, G.S. Camarda, Y. Cui, R.B. James, Growth interface study of CdTeSe crystals grown by the THM technique, J. Cryst. Growth. 616 (2023) 127261. https://doi.org/10.1016/j.jcrysgro.2023.127261.

[39] T.E. Schlesinger, J.E. Toney, H. Yoon, E.Y. Lee, B.A. Brunett, L. Franks, R.B. James, Cadmium zinc telluride and its use as a nuclear radiation detector material, Mater. Sci. Eng. R Reports. 32 (2001) 103–189. https://doi.org/10.1016/S0927-796X(01)00027-4.

[40] A. Mycielski, A. Burger, M. Sowinska, M. Groza, A. Szadkowski, P. Wojnar, B. Witkowska, W.

Kaliszek, P. Siffert, Is the (Cd,Mn)Te crystal a prospective material for X-ray and γ-ray detectors?, Phys. Status Solidi C. 2 (2005) 1578–1585.

[41] A. Wardak, D.M. Kochanowska, M. Kochański, M. Dopierała, A. Sulich, J. Gdański, A. Marciniak, A. Mycielski, Effect of doping and annealing on resistivity, mobility-lifetime product, and detector response of (Cd,Mn)Te, J. Alloys Compd. 936 (2023) 168280. https://doi.org/10.1016/j.jallcom.2022.168280.

[42] U.N. Roy, A. Burger, R.B. James, Growth of CdZnTe crystals by the traveling heater method, J. Cryst. Growth. 379 (2013) 57–62. https://doi.org/10.1016/j.jcrysgro.2012.11.047.

[43] U.N. Roy, G.S. Camarda, Y. Cui, R. Gul, G. Yang, J. Zazvorka, V. Dedic, J. Franc, R.B. James, Evaluation of CdZnTeSe as a high-quality gamma-ray spectroscopic material with better compositional homogeneity and reduced defects, Sci. Rep. 9 (2019) 7303. https://doi.org/10.1038/s41598-019-43778-3.

[44] U.N. Roy, G.S. Camarda, Y. Cui, R.B. James, X-ray topographic study of Bridgman-grown CdZnTeSe, J. Cryst. Growth. 546 (2020) 125753.

[45] K. Kim, Y. Kim, J. Franc, P. Fochuk, A.E. Bolotnikov, R.B. James, Enhanced hole mobility-lifetime product in selenium-added CdTe compounds, Nucl. Instruments Methods Phys. Res. Sect. A Accel. Spectrometers, Detect. Assoc. Equip. 1053 (2023) 168363. https://doi.org/10.1016/j.nima.2023.168363.

[46] J. Zhang, W. Jie, T. Wang, D. Zeng, S. Ma, H. Hua, B. Yang, Crystal growth and characterization of Cd0.8Mn0.2Te using Vertical Bridgman method, Mater. Res. Bull. 43 (2008) 1239–1245. https://doi.org/10.1016/j.materresbull.2007.05.029.

[47] A. Tanaka, Y. Masa, S. Seto, T. Kawasaki, Zinc and selenium co-doped CdTe substrates lattice matched to HgCdTe, J. Cryst. Growth. 94 (1989) 166–170. https://doi.org/https://doi.org/10.1016/0022-0248(89)90615-5.

[48] J. González-Hernández, E. López-Cruz, D.D. Allred, W.P. Allred, Photoluminescence studies in ZnxCd1−xTe single crystals, J. Vac. Sci. Technol. A Vacuum, Surfaces, Film. 8 (1990) 3255–

3259. https://doi.org/10.1116/1.576574.

[49] Y.R. Lee, A.K. Ramdas, A piezomodulation study of the absorption edge and Mn++ internal transition in Cd1−xMnxTe, a prototype of diluted magnetic semiconductors, Solid State Commun. 51 (1984) 861–863. https://doi.org/https://doi.org/10.1016/0038-1098(84)91088-3.

[50] L. Hannachi, N. Bouarissa, Electronic structure and optical properties of CdSexTe1−x mixed crystals, Superlattices Microstruct. 44 (2008) 794–801.

[51] P.D. Brown, K. Durose, G.J. Russell, J. Woods, The absolute determination of CdTe crystal polarity, J. Cryst. Growth. 101 (1990) 211–215. https://doi.org/https://doi.org/10.1016/0022-0248(90)90968-Q.

[52] M. Inoue, I. Teramoto, S. Takayanagi, Etch pits and polarity in CdTe crystals, J. Appl. Phys. 33 (1962) 2578–2582. https://doi.org/10.1063/1.1729023.

[53] J. Franc, P. Moravec, V. Dědič, U. Roy, H. Elhadidy, P. Minárik, V. Šíma, Microhardness study of Cd1-x ZnxTe1-ySey crystals for X-ray and gamma ray detectors, Mater. Today Commun. 24 (2020) 101014.

[54] K. Guergouri, R. Triboulet, A. Tromson-Carli, Y. Marfaing, Solution hardening and dislocation density reduction in CdTe crystals by Zn addition, J. Cryst. Growth. 86 (1988) 61–65.

[55] L. Marchini, A. Zappettini, M. Zha, N. Zambelli, A.E. Bolotnikov, G.S. Camarda, R.B. James, Crystal defects in CdZnTe crystals grown by the modified low-pressure Bridgman method, IEEE Trans. Nucl. Sci. 59 (2012) 264–267. https://doi.org/10.1109/TNS.2011.2181414.

[56] A.E. Bolotnikov, G.S. Camarda, Y. Cui, G. Yang, A. Hossain, K. Kim, R.B. James, Characterization and evaluation of extended defects in CZT crystals for gamma-ray detectors, J. Cryst. Growth. 379 (2013) 46–56. https://doi.org/10.1016/j.jcrysgro.2013.01.048.

[57] J. Butcher, M. Hamade, M. Petryk, A.E. Bolotnikov, G.S. Camarda, Y. Cui, G. De Geronimo, J. Fried, A. Hossain, K.H. Kim, E. Vernon, G. Yang, R.B. James, Drift time variations in CdZnTe detectors measured with alpha particles and gamma rays: Their correlation with detector response, IEEE Trans. Nucl. Sci. 60 (2013) 1189–1196. https://doi.org/10.1109/TNS.2012.2234762.

[58] U.N. Roy, G.S. Camarda, Y. Cui, R.B. James, Characterization of large-volume Frisch grid detector fabricated from as-grown CdZnTeSe, Appl. Phys. Lett. 115 (2019) 242102. https://doi.org/10.1063/1.5133389.

[59] J. Zou, A. Fauler, A.S. Senchenkov, N.N. Kolesnikov, L. Kirste, M.P. Kabukcuoglu, E. Hamann, A. Cecilia, M. Fiederle, Characterization of structural defects in (Cd,Zn)Te crystals grown by the travelling heater method, Crystals. 11 (2021) 1402. https://doi.org/10.3390/cryst11111402.

[60] K. Nakagawa, K. Maeda, S. Takeuchi, Observation of dislocations in cadmium telluride by cathodoluminescence microscopy, Appl. Phys. Lett. 34 (1979) 574–575. https://doi.org/10.1063/1.90871.

[61] L. Xu, B. Yu, G. Yu, H. Liu, L. Zhang, X. Li, P. Huang, B. Wang, S. Wang, Study on the morphology of dislocation-related etch pits on pyramidal faces of KDP crystals, CrystEngComm. 23 (2021) 2556–2562. https://doi.org/10.1039/d1ce00069a.

[62] C.G. Darwin, The reflexion of X-rays from imperfect crystals, London, Edinburgh, Dublin Philos. Mag. J. Sci. 43 (1922) 800–829. https://doi.org/10.1080/14786442208633940.

[63] U.N. Roy, J.N. Baker, G.S. Camarda, Y. Cui, G. Yang, R.B. James, Evaluation of crystalline quality of traveling heater method (THM) grown Cd0.9Zn0.1Te0.98Se0.02 crystals, Appl. Phys. Lett. 120 (2022) 242103. https://doi.org/10.1063/5.0093255.

[64] G.A. Carini, G.S. Camarda, Z. Zhong, D.P. Siddons, A.E. Bolotnikov, G.W. Wright, B. Barber, C. Arnone, R.B. James, High-energy X-ray diffraction and topography investigation of CdZnTe, J. Electron. Mater. 34 (2005) 804–810. https://doi.org/10.1007/s11664-005-0024-6.

[65] D. Zeng, W. Jie, T. Wang, G. Zha, J. Zhang, Effects of mosaic structure on the physical properties of CdZnTe crystals, Nucl. Instruments Methods Phys. Res. A. 586 (2008) 439–443. https://doi.org/10.1016/j.nima.2007.12.033.

[66] V. Carcelén, K.H. Kim, G.S. Camarda, A.E. Bolotnikov, A. Hossain, G. Yang, J. Crocco, H. Bensalah, F. Dierre, E. Diéguez, R.B. James, Pt coldfinger improves quality of Bridgman-grown Cd 0.9Zn0.1Te:Bi crystals, J. Cryst. Growth. 338 (2012) 1–5.

https://doi.org/10.1016/j.jcrysgro.2011.09.031.

[67] P. Yu, Y. Chen, W. Li, W. Liu, B. Liu, J. Yang, K. Ni, L. Luan, J. Zheng, Z. Li, M. Bai, G. Sun, H. Li, W. Jie, Study of detector-grade CdMnTe:In crystals obtained by a multi-step post-growth annealing method, Crystals. 8 (2018) 387. https://doi.org/10.3390/cryst8100387.

[68] P. Yu, Y. Xu, L. Luan, Y. Du, J. Zheng, H. Li, W. Jie, Quality improvement of CdMnTe:In single crystals by an effective post-growth annealing, J. Cryst. Growth. 451 (2016) 194–199. https://doi.org/10.1016/j.jcrysgro.2016.07.009.

[69] D. Kochanowska, M. Rasiński, M. Witkowska-Baran, M. Lewandowska, A. Mycielski, Studies of the surface regions of (Cd, Mn) Te crystals, Phys. Status Solidi. 11 (2014) 1523–1527. https://doi.org/https://doi.org/10.1002/pssc.201300711.

[70] A. Wardak, W. Chromiński, A. Reszka, D. Kochanowska, M. Witkowska-Baran, M. Lewandowska, A. Mycielski, Stresses caused by Cd and Te inclusions in CdMnTe crystals and their impact on charge carrier transport, J. Alloys Compd. 874 (2021) 159941. https://doi.org/https://doi.org/10.1016/j.jallcom.2021.159941.

[71] M. Witkowska-Baran, D.M. Kochanowska, A. Mycielski, R. Jakieła, A. Wittlin, W. Knoff, A. Suchocki, P. Nowakowski, K. Korona, Influence of annealing on the properties of (Cd,Mn)Te crystals, Phys. Status Solidi Curr. Top. Solid State Phys. 11 (2014) 1528–1532. https://doi.org/10.1002/pssc.201300746.

[72] B.K. Meyer, P. Omling, E. Weigel, G. Müller-Vogt, F center in CdTe, Phys. Rev. B. 46 (1992) 15135.

[73] J.H. Yang, W.J. Yin, J.S. Park, J. Ma, S.H. Wei, Review on first-principles study of defect properties of CdTe as a solar cell absorber, Semicond. Sci. Technol. 31 (2016) 083002. https://doi.org/10.1088/0268-1242/31/8/083002.